\documentclass[twocolumn]{aastex631}
\usepackage{listings}
\usepackage{hyperref}
\usepackage{threeparttable}

\shorttitle{CHIPS-Ibc}
\shortauthors{Takei et al.}

\begin{document}

\title{Simulating Hydrogen-poor Interaction-Powered Supernovae with CHIPS}

\author{Yuki Takei}
\affiliation{Research Center for the Early Universe (RESCEU), Graduate School of Science, The University of Tokyo, 7-3-1 Hongo, Bunkyo-ku, Tokyo 113-0033, Japan}
\affiliation{Astrophysical Big Bang Laboratory, RIKEN, 2-1 Hirosawa, Wako, Saitama 351-0198, Japan}

\author{Daichi Tsuna}
\affiliation{TAPIR, Mailcode 350-17, California Institute of Technology, Pasadena, CA 91125, USA}
\affiliation{Research Center for the Early Universe (RESCEU), Graduate School of Science, The University of Tokyo, 7-3-1 Hongo, Bunkyo-ku, Tokyo 113-0033, Japan}

\author{Takatoshi Ko}
\affiliation{Research Center for the Early Universe (RESCEU), Graduate School of Science, The University of Tokyo, 7-3-1 Hongo, Bunkyo-ku, Tokyo 113-0033, Japan}
\affiliation{Department of Astronomy, Graduate School of Science, The University of Tokyo, Tokyo, Japan}
\affiliation{Astrophysical Big Bang Laboratory, RIKEN, 2-1 Hirosawa, Wako, Saitama 351-0198, Japan}

\author{Toshikazu Shigeyama}
\affiliation{Research Center for the Early Universe (RESCEU), Graduate School of Science, The University of Tokyo, 7-3-1 Hongo, Bunkyo-ku, Tokyo 113-0033, Japan}
\affiliation{Department of Astronomy, Graduate School of Science, The University of Tokyo, Tokyo, Japan}

\correspondingauthor{Yuki Takei}
\email{takei@resceu.s.u-tokyo.ac.jp}

%% Note that the \and command from previous versions of AASTeX is now
%% depreciated in this version as it is no longer necessary. AASTeX 
%% automatically takes care of all commas and "and"s between authors names.

%% AASTeX 6.31 has the new \collaboration and \nocollaboration commands to
%% provide the collaboration status of a group of authors. These commands 
%% can be used either before or after the list of corresponding authors. The
%% argument for \collaboration is the collaboration identifier. Authors are
%% encouraged to surround collaboration identifiers with ()s. The 
%% \nocollaboration command takes no argument and exists to indicate that
%% the nearby authors are not part of surrounding collaborations.

%% Mark off the abstract in the ``abstract'' environment. 
\begin{abstract}
We present the updated open-source code Complete History of Interaction-Powered Supernovae (\texttt{CHIPS}) that can be applied to modeling supernovae (SNe) arising from an interaction with massive circumstellar medium (CSM) as well as the formation process of the CSM. Our update mainly concerns with extensions to hydrogen-poor SNe from stripped progenitors, targeting modeling of interaction-powered SNe Ibc such as Type Ibn and Icn SNe. We successfully reproduce the basic properties of the light curves of these types of SNe that occur after partial eruption of the outermost layer with a mass of $0.01$--$0.1\,M_\odot$ at $\lesssim 1$ year before explosion. We also find that the luminosity of the observed precursors can be naturally explained by the outburst that creates the dense CSM, given that the energy of the outburst is efficiently dissipated by collision with an external material, possibly generated by a previous mass eruption. We discuss possible scenarios causing eruptive mass-loss based on our results.
\end{abstract}

%% Keywords should appear after the \end{abstract} command. 
%% The AAS Journals now uses Unified Astronomy Thesaurus concepts:
%% https://astrothesaurus.org
%% You will be asked to selected these concepts during the submission process
%% but this old "keyword" functionality is maintained in case authors want
%% to include these concepts in their preprints.

\section{Introduction}
Supernovae (SNe) are classified in various types, reflecting the diverse evolution pathways of the progenitors towards core-collapse. Explosion of a massive star inside a dense, hydrogen-rich circumstellar medium (CSM) results in a supernova categorized to a luminous subclass of SNe called Type II``n", indicating narrow emission lines in the spectrum. These SNe were first identified by \citet{1990MNRAS.244..269S} \citep[see also][for a review]{Filippenko_et_al_1997}, and are often powered by the interaction between the fast SN ejecta and slowly-moving dense CSM. This is in contrast to other ``normal" SNe, which are mainly powered by the internal energy and/or the radioactive decay of $^{56}\mathrm{Ni}$ and $^{56}\mathrm{Co}$ in the ejecta. 

Interestingly, observations of such SNe have confirmed that their progenitors experience mass loss orders of magnitude stronger than the observed local supergiants \citep[e.g.,][]{DeJager88,vanLoon05,Mauron11,Beasor20}.
Furthermore, it is suggested that the early-phase emission from SNe II indicate a large majority of their progenitors having dense CSM in the vicinities, which means that the progenitors selectively lose a significant amount of the envelopes near the end of their lives \citep{Khazov_2016,Yaron_2017,Morozova18,Foerster_et_al_2018,Bruch_2021}.
Although no direct observational evidence of any mass loss mechanism exist so far, possible scenarios are gravity waves excited by the sudden increase of nuclear burning energy \citep[e.g.,][]{2012MNRAS.423L..92Q,Shiode_Quataert_2014,Fuller_2017,Fuller_Ro_2018,Leung_et_al_2021,Wu_Fuller_2022}, pulsational pair instability \citep{Woosley_2017,Woosley2019,Leung19,Renzo_et_al_2020}, or a binary interaction \citep{Chevalier_2012,Metzger_2022,Wu22_Ibc}. 

Thanks to recent high cadence surveys, more rapidly evolving SNe Ibn have been found, which have prominent narrow helium lines but weak or no hydrogen lines in their spectra \citep[e.g.,][]{Matheson_et_al_2000, Pastorello_et_al_2007,Hosseinzadeh_et_al_2017}. SN Ibn is a rarer class of Core-collapse (CC) SNe \citep[with an event rate of $\sim1\%$ of CCSNe,][]{Pastorello08, Maeda_Moriya_2022} compared to Type IIn which accounts for $\sim5$--$10\%$ of all CCSNe \citep{Smith_et_al_2011,Graur17,Cold23}. The light curve (LC) around peak is thought to be mainly powered by interaction between ejecta and helium-rich CSM. 

Originally, hydrogen-deficient (Type Ibc) SNe are thought to be an explosion of a Wolf-Rayet (WR) star, losing its hydrogen-rich envelope due to the strong stellar wind or binary interaction \citep{1990ApJ...361L..23S,Heger_et_al_2003,Crowther_2007,Smith_et_al_2011}. The important implication arising from the observation of SNe Ibc is that the progenitor could be sometimes a helium star whose mass is lower than single WR stars \citep[with masses of typically 10--25$\,M_\odot$,][]{Crowther_2007}, which strongly suggests that the progenitor is in a close binary system \citep[e.g.,][]{Lyman_et_al_2016,Hosseinzadeh_et_al_2019,Dessart_et_al_2022,Wang_et_al_2023}. However, the detailed process(es) leading to the generation of the dense CSM is still poorly known.

Following the first detection of the precursor of the Type Ibn SN 2006jc by an amateur astronomer Koichi Itagaki \citep{Nakano_et_al_2006}, precursors of these SNe are starting to be observed thanks to high-cadence pre-explosion imaging \citep[e.g.,][]{Ofek14,Margutti_et_al_2014,Strotjohann_et_al_2021,Jacobson_et_al_2022,Hiramatsu23_2021qqp}.
Since the luminosity of the precursors exceeding the Eddington limit of a star with a mass of 10\,$M_\odot$ indicates an eruptive mass loss, the property of the precursors might be a direct probe of these mass-loss events. Thus, modeling LCs of interacting SNe including eruptions from a progenitor still plays a crucial role in extracting the properties (e.g., mass, extent, density profile) of the CSM and probing its origin. Though precursor events have been detected for a few SNe Ibn at present because of too faint precursors, the importance of modeling is growing, as the number of observations of interacting SNe is expected to dramatically increase with upcoming optical and infrared surveys such as Legacy Survey of Space and Time by Rubin Observatory \citep[LSST,][]{LSST_2019} and Roman Space Telescope \citep{Spergel_et_al_2015}.

Recently, \citet{Takei_et_al_2022} has developed an open-source code, Complete History of Interaction-Powered Supernovae (\texttt{CHIPS}), that simulates the creation of the dense CSM as well as the subsequent SN explosion. This code enabled a novel framework to obtain the CSM profile itself in a self-consistent manner, rather than putting the CSM profile by hand as usually done in conventional modelings of interacting SNe. Although \texttt{CHIPS} has been verified to successfully reproduce the photometric observations of SNe IIn, the progenitor models were limited to red supergiants (RSGs) with an extended hydrogen-rich envelope. For modeling SNe Ibc, whose progenitors are stripped stars having completely different radii and composition from RSGs, several updates to the physics in the code were required for both modeling the dense CSM and the subsequent explosion. Furthermore, various observational findings on these SNe have been reported since the previous release of \texttt{CHIPS}, which is worth comparing with our updated modeling.

In this paper, we introduce the updated \texttt{CHIPS} code\footnote{The updated code is labeled as Version 2, archived in Zenodo \citep{CHIPS}} , including the newly implemented options, that can be applied to modeling of Type Ibc SNe with CSM interaction. We show that one can model with this code various properties of interacting SNe Ibc, such as the properties of the CSM, light curves of SN precursors, and the light curves of the final SN. By using \texttt{CHIPS}, we also aim to constrain the properties of the progenitor.

This paper is organized as follows; In Section \ref{sec:technical_details}, we describe some updates on the \texttt{CHIPS} code. In Section \ref{sec:Results}, we present our results with the updated \texttt{CHIPS}, and compare the results with the observed Ibn SNe and precursors. Finally, Section \ref{sec:conclusion} provides the summary and discussion.

\begin{table*}
\centering
\begin{threeparttable}
\caption{The key parameters in the CHIPS code that the user can tune. Progenitor models of ZAMS mass $15$--$29$\,$M_\odot$ have been prepared for this paper, but the code can use as input any \texttt{MESA} model supplied by the user. In the last column we show the range of values recommended by the authors, where the code generally runs stably and reproduce the characteristics of the ejecta and CSM observed in interacting Type Ibc SNe.}
\begin{tabular}{c|cc}
\hline\hline
Parameter & Definition & Recommended values \\ \hline
$M_\mathrm{ZAMS}$ & Mass of star at ZAMS ($15$--$29$\,$M_\odot$ prepared in this paper) & --\\
$f_{\rm inj}$ & Injected energy nomalized by the envelope's binding energy & $0.3$--$0.8$\\
$t_{\rm inj}$ & Time from energy injection to explosion & $0.1$--$1$\,yr\\
$E_{\rm ej}$ & Explosion energy of the SN ejecta & $10^{50}$--$10^{52}$ erg\\
$M_{\rm Ni}$ & The mass of $^{56}$Ni in the SN ejecta & $0$--$1\,M_\odot$ \\ \hline
\end{tabular}
\label{table:Parameters}
\end{threeparttable}
\end{table*}

\section{Updates on the CHIPS Code}
\label{sec:technical_details}
In this section we describe updates and modifications to \texttt{CHIPS} from \citet{Takei_et_al_2022} that only covered RSGs, to extend it to apply to stripped stars and make it compatible with interaction-powered Type Ibc SNe. We refer to \citet{Takei_et_al_2022} for details of the code structure and the example output files, but make a brief summary here for completeness \footnote{The documentation of \texttt{CHIPS} is also available on \href{https://github.com/DTsuna/CHIPS}{https://github.com/DTsuna/CHIPS} upon acceptance of this manuscript.}.

The code is composed of two main numerical calculations, named as ``eruption" and ``LC". \texttt{CHIPS} starts with reading a progenitor model generated with \texttt{MESA}. The user can use the sample stellar models we have generated, or prepare their own \texttt{MESA} model. In the eruption part, the mass eruption from a progenitor is simulated using a radiation hydrodynamics code developed in \citet{Kuriyama20a}. The resulting CSM is the input for the LC part, where bolometric and multi-band light curves can be calculated using the code developed in \citet{Takei20}.

In Table \ref{table:Parameters} we list the five key parameters in the \texttt{CHIPS} code, $M_\mathrm{ZAMS},\,f_\mathrm{inj},\,t_\mathrm{inj},\,E_{\rm ej},\,M_\mathrm{Ni}$, with recommended values. $M_\mathrm{ZAMS}$ denotes the mass of star at Zero-Age Main Sequence (ZAMS), used in Section \ref{sec:progenitor_models}. Given a progenitor model, the code injects internal energy at the inner edge of an arbitrarily chosen computational region, by default from the base of the outermost envelope layer to the surface. The energy is specified by a parameter $f_\mathrm{inj}$ that is scaled with the envelope's binding energy (Section \ref{sec:mass_eruption_explanation}). The time from energy injection to core collapse is set by a parameter $t_\mathrm{inj}$ (Section \ref{sec:construction_of_CSM}), that mainly controls the CSM extent and hence the duration of the CSM interaction. Finally, $E_{\rm ej}$ and $M_\mathrm{Ni}$ are respectively the kinetic energy of the ejecta and mass of radioactive $^{56}$Ni resultant from a core collapse (Section \ref{sec:56ni_co}).

\subsection{New Sample Progenitors}
\label{sec:progenitor_models}
\begin{figure*}
 \centering
 \includegraphics[width=\linewidth]{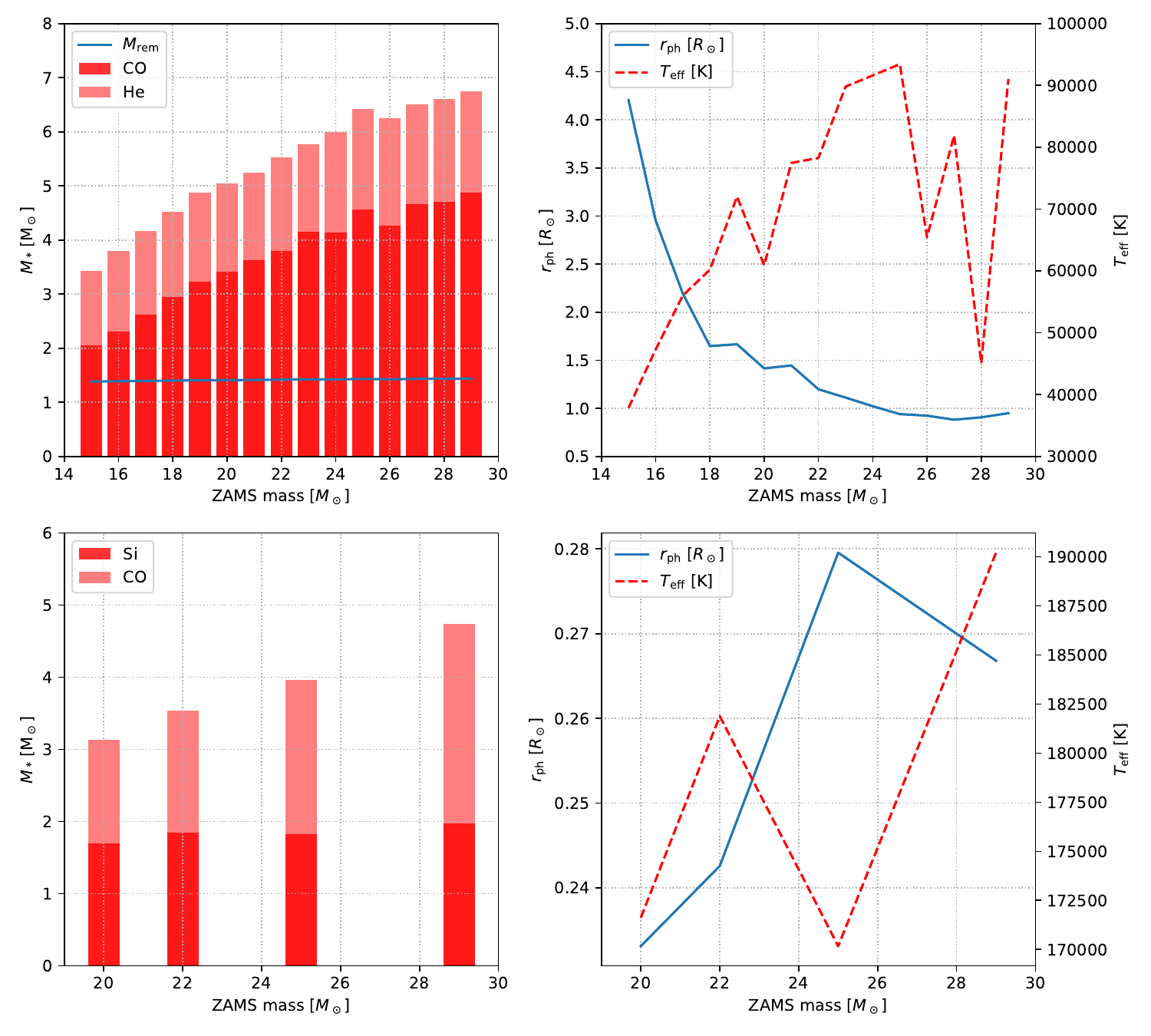}
\caption{Properties of the progenitors we adopt in this work. The left panels show the masses of the helium core and the carbon-oxygen core for progenitors of SNe Ibn (top), the masses of the carbon-oxygen core and silicon core for progenitors of SNe Icn (bottom), and the remnant mass $M_{\rm rem}$ obtained from equation (\ref{eq:remnant_mass}). The right panels show photospheric radius (in $R_\odot$) of each star and the effective temperature (in Kelvins). These progenitors are available (stored in \texttt{input/mesa\_models\_ibn}, \texttt{input/mesa\_models\_icn}) when downloading the \texttt{CHIPS} code, and one can use them without running the \texttt{MESA} calculation. }
 \label{fig:progenitors}
 \end{figure*}

For our code we prepared additional sample helium star and helium-poor star models using the 1D stellar evolution code \verb|MESA| version 12778 \citep{Paxton11,Paxton13,Paxton15,Paxton18,Paxton19,Jermyn23}\footnote{The sample inlist files are available on Zenodo under the following DOI: \url{https://doi.org/10.5281/zenodo.10052341}}. The resulting photospheric radii, effective temperatures, total masses, and mass of each layer at core collapse are summarized in Figure \ref{fig:progenitors}.
 
We prepare 15 helium star models with ZAMS masses of $15,\,16,\,\cdots,\,29\,M_\odot$ and the metallicity of $Z=Z_\odot=0.014$ basically using the \texttt{example\_make\_pre\_ccsn} test suite in \texttt{MESA}.
For the mass-loss we have adopted the standard ``Dutch" wind scheme for both hot and cold winds, with the \texttt{Dutch\_scaling\_factor} set to 1. Convective mixing is included by mixing length theory. We use the mixing-length parameter $\alpha_\mathrm{MLT}=3$ if the hydrogen fraction in a cell is $\geq0.5$, and $\alpha_\mathrm{MLT}=1.5$ otherwise.
In order to create helium stars, we remove the hydrogen-rich envelope in the middle of the evolution. After the hydrogen burning stage when the core hydrogen mass fraction drops to less than $10^{-6}$, the hydrogen envelope is removed with a constant rate of $10^{-4}\,M_\odot\,{\rm yr^{-1}}$ for models with initial masses of $M_\mathrm{ZAMS}=15,\,16,\,\cdots,\,23\,M_\odot$, which are done by setting the parameter \texttt{mass\_change} in the file \texttt{inlist\_remove} to \texttt{-1d-4}. For other models with larger masses, this parameter is enhanced to \texttt{-2d-4} to avoid the severe condition of the time step. We continue the stripping until $0.01\,M_\odot$ of hydrogen remains, which is quickly removed by the subsequent wind mass-loss.

We generate four more samples of helium-poor stars, which can be used as progenitors of Type Ic SNe with signatures of interaction. The ZAMS mass for these models are $M_{\rm ZAMS}=20,\, 22,\,25,\,29\,M_\odot$. For these models, after we strip the hydrogen envelope by the same process as above, we further remove the helium envelope just after core helium depletion when its mass fraction drops to less than $10^{-2}$. For this second stage of stripping we adopt a stronger mass-loss rate of $5\times 10^{-4}\,M_\odot\ $yr$^{-1}$, because of the faster evolution of the stellar core than the first stage. We stop the stripping when $0.03\,M_\odot$ of helium remains, and evolve the star until core-collapse. For all four progenitors at core-collapse, the mass of leftover helium in the outermost CO layer is $\lesssim 0.003\,M_\odot$, which is low enough to be compatible with the upper limits suggested for a stripped-envelope SNe to be classified as Type Ic \citep[e.g.,][]{Hachinger12,Teffs20,Williamson21}.

\subsection{Updates to Mass Eruption Calculation}
\subsubsection{Opacity in the Eruption Code}
\label{sec:opacity}
We simulate eruptive mass-loss by solving one-dimensional Lagrangian radiation hydrodynamics equations. The opacity model in the eruption calculations \citep{Kuriyama20a}, adopted in the earlier release of \texttt{CHIPS}, is an analytical formula that is calibrated only for hydrogen-rich gas. The formula inevitably introduces a finite contribution from negative hydrogen ion even for hydrogen-free gas, which should be absent in the envelope of hydrogen-poor progenitors. A more rigorous formulation of the opacity is desirable, especially when we consider mass eruption from such stars. This can also be important for modeling the precursors (Section \ref{sec:precursor}), where accurate inputs for radiative transfer are important (see also \citealt{Tsuna_et_al_2023}).

We thus added a new argument \verb|--opacity-table|, that enables the user to add a specific opacity table as a function of temperature and density. The table should be rectangular, with rows of temperature in [K] and columns of $R=\rho/(10^{-6}T)^{3}$ (in cgs units). Such format is adopted for commonly used opacity tables such as OPAL \citep{Ross} and \AE SOPUS \citep{Marigo_Aringer_2009,Marigo_et_al_2022}. 

For the results shown in this work we use the tabulated Rosseland-mean opacities, OPAL \citep{Ross} and \AE SOPUS 2.0 \citep{Marigo_Aringer_2009,Marigo_et_al_2022}\footnote{We retrieved the OPAL opacity table data from \href{https://opalopacity.llnl.gov/opal.html}{https://opalopacity.llnl.gov/opal.html}, and the \AE SOPUS 2.0 table data from \href{http://stev.oapd.inaf.it/cgi-bin/aesopus}{http://stev.oapd.inaf.it/cgi-bin/aesopus}. The sample tables are in the directory \texttt{input/rosseland}.}. The OPAL opacity table covers the temperature range of $3.75\leq\log T\,[{\rm K}] \leq8.70$, while \AE SOPUS $3.2\leq\log T\,[{\rm K}]\leq4.50$. For the temperature range that overlap, we adopt the table by \AE SOPUS for $\log T\leq4.0$, and that by OPAL for $\log T>4.0$. 
The density range covered by both of the tables is $-8.0\leq\log R\leq1.0$. If $T$ or $R$ is out of boundaries, we use the edge values.

\subsubsection{Energy Injection and Simulation Grid for Eruption}
\label{sec:mass_eruption_explanation}
We make a few updates on how to inject the thermal energy into the base of the layer because stripped stars usually have a radius of $\sim R_\odot$ much smaller than RSGs, except for an extreme case, i.e., low-mass progenitor. The inner boundary is set to $0.2\,M_\odot$ outside the carbon-oxygen (silicon) core for progenitors of Type Ibn (Icn). 

The duration of energy injection (in seconds) can be specified by the user by the argument \verb|--injection-duration|, with a default value of 1000 seconds. This default value is arbitrary, and should depend on the progenitor, as well as the mechanism that triggers the eruption. \cite{Ko_et_al_2022} found from a parameter study of both the duration and $f_{\rm inj}$ that mass loss can be significantly suppressed for durations comparable to the dynamical timescale at the envelope 
 \begin{eqnarray}
 t_{\rm dyn} \sim \left(\frac{R_*^3}{GM_*}\right)^{1/2} \sim 2.6\times10^3\,{\rm s} \left(\frac{R_*}{2\,R_\odot}\right)^{3/2}\left(\frac{M_*}{3\,M_\odot}\right)^{-1/2},
 \end{eqnarray}
where $G,\,M_*,\,R_*$ denote the gravitational constant, the progenitor mass, and the progenitor radius respectively. However, the mass loss was insensitive to the duration when it was much less than $t_{\rm dyn}$. Furthermore, the net effect of raising the duration comparable to $t_{\rm dyn}$ is nearly equivalent to adopting a lower $f_{\rm inj}$.

For the purpose of this paper, in order to simplify the parameter space we fix the duration to 10 seconds, which is much shorter than the dynamical timescale in the helium layer. We note that for a fixed $f_{\rm inj}$, this thus gives an upper limit of mass that can be erupted. We discuss in detail the dependence of the CSM on the duration in Appendix \ref{sec:dependence_duration}.

The mass coordinates of the cells can either be uniform spacing, or logarithmic spacing with finer meshing outside with an optical depth of $0.1(\kappa/1\ {\rm cm^2\ g^{-1}} )$ in the outermost cell. For radiation hydrodynamical simulations with helium stars, the latter should be chosen to fully resolve the photosphere close to stellar surface. To save the computational cost, the number of cells, defined by the argument \verb|hydroNumMesh|, is reduced to 4,000 from the default value of 10,000. We have confirmed with several models that reducing the resolution to this value does not change the resultant profiles of the CSM. 

\subsubsection{Constructing the CSM Profile at Core-collapse}
\label{sec:construction_of_CSM}

\texttt{CHIPS} evolves the CSM until $t=2t_\mathrm{dyn}$ using radiation hydrodynamical simulations, when the mass of the unbound CSM is well converged (see also Appendix \ref{sec:dependence_duration}). While this is about a year for RSG progenitors, for stripped stars this is about an hour, much shorter than the typical CSM expansion timescales inferred from SNe Ibn/Icn (months to years). Simulating the entire computational region for such long timescales is computationally infeasible, due to the much smaller timesteps required to follow the inner part of the region. 

Fortunately, the magnitude of the pressure gradient in the erupted CSM is significantly smaller than the gravity at a few $t_{\rm dyn}$, which greatly simplifies the dynamical evolution of the CSM \citep{Kuriyama20a}. If thermal pressure is negligible, each computational cell obeys the following equation of motion,
 \begin{eqnarray}
     \frac{d^{2}r}{dt^{2}}=-\frac{GM_r}{r^{2}},
     \label{eq:eq_of_motion}
 \end{eqnarray}
where $M_r$ denotes the enclosed mass within radius $r$. Thus, for the unbound part of the CSM we stop the simulations at $t=2t_\mathrm{dyn}$ and solve the trajectory of each fluid element independently with equation (\ref{eq:eq_of_motion}).  We integrate the above equation from $t=2t_\mathrm{dyn}$ to $t=t_\mathrm{inj}$ with a Runge-Kutta method in scipy (\verb|integrate.rk45|). 

For the bound part of the CSM, most of the fluid elements would fall towards the star before $t=t_\mathrm{inj}$. If feedback is weak from the leftover star, a small amount of mass would be still falling back, forming a characteristic density profile of $\rho_{\rm CSM}\propto r^{-1.5}$ near the surface \citep{Tsuna21b,Tsuna_Takei_2023}. These works have also found that the CSM density profile can be well fitted by a double power-law
\begin{eqnarray}
    \rho_\mathrm{CSM}(r)=\rho_*\left[\frac{(r/r_*)^{1.5/y_*}+(r/r_*)^{n_\mathrm{out}/y_*}}{2}\right]^{-y_*},
    \label{eq:rho_CSM_analytical}
\end{eqnarray}
where $r_*,\,\rho_*,\,n_\mathrm{out}$ are the fitting parameters, and $y_*$ is set to 2.5 for all cases\footnote{This parameter sets only the curvature of the profile around $r_*$, and its choice does not largely affect the CSM density profile, as discussed in \citet{Tsuna21b}.}. 
We use this formula to fit the unbound part of the CSM obtained above and extend the CSM to the surface of the progenitor.
The outer boundary of the computational region $r_\mathrm{out}$ is set to $3\times10^{16}\,{\rm cm}$. In the usual case that the outer edge of the CSM is smaller than $r_\mathrm{out}$, the CSM profile is smoothly transitioned to a steady wind profile with $\rho\propto r^{-2}$. The wind velocity $v_\mathrm{w}$ is approximated by the escape velocity at the stellar surface, $(2GM_*/R_*)^{1/2}$. Then the mass-loss rate is written as $\dot{M}=4\pi v_\mathrm{w}\rho_{n}r_{n}^{2}$, where $\rho_{n},\,r_{n}$ denote the density and radius at the outermost cell of the computational region.
The detailed choices of mass-loss rate and wind velocity at this outer region do not affect the resulting LCs.

We note that this density profile characterized by fallback can be altered at late times, due to feedback from stellar wind and radiation from the leftover progenitor. \cite{Tsuna_Takei_2023} found that depending on the progenitor and the CSM mass, these feedback processes can affect the fallback for $t_{\rm inj}$ of several years or longer. Strong feedback creates a CSM detached from the star, consistent with what is seen in some Type Ic SNe with late-time CSM interaction such as SN2021ocs and SN2022xxf \citep{Kuncarayakti22,Kuncarayakti_2023}. In summary, we advise that the user should be cautious of the validity of this profile when modeling light curves with $t_{\rm inj}\gg 1$ yr.

 \subsubsection{Light Curves of the Outbursts}
 \label{sec:precursor}
When the shock formed after energy injection travels to the surface, it would disrupt the outer part of the star and cause an outburst. Such outburst may accompany detectable brightening of the star (i.e. SN precursor), which may enable us to put a constraint on $t_\mathrm{inj}$ and $f_\mathrm{inj}$ \citep[e.g.,][]{Strotjohann_et_al_2021,Jacobson_et_al_2022,Tsuna_et_al_2023}. 

To this end, our code also records the information of the fluid at the photosphere at every timestep, such as the luminosity, radii, density, and temperature, as a file \verb|photosphere.txt| in the directory \verb|EruptionFiles|. Such output was used in \cite{Tsuna_et_al_2023} when modeling precursors from RSGs, and in Section \ref{sec:precursor} we use this to construct light curves for the outburst of hydrogen-poor stars.

\subsection{Updates to SN Light Curve Calculation}
\label{sec:light_curve}
The SN light curve calculation code is composed of two parts. First, the code solves the dynamics of the shocks formed by the collision between the SN ejecta and the CSM, resolving the shocked region assuming steady state at the rest frame of each shock. This code outputs the radiative flux at the forward shock, which is then used  for the next step that calculates radiative transfer and energy equations in the unshocked CSM to obtain the observed luminosity. Here we describe updates to the LC part, mainly due to address the issues arising from changes in the properties of the progenitor. We also explain how we implement thermal emission from the radioactive decay of $^{56}$Ni and $^{56}$Co, which can be important for hydrogen-poor SNe even if they are powered by interaction with the CSM around the luminosity peak.

\subsubsection{Parameter setup for the SN explosion}\label{sec:parameter}
We do not solve the SN explosion itself, but generate a density and velocity profile of the homologous SN ejecta following the model of \cite{Matzner99}. The density profile follows a double power-law of velocity, with the inner power-law index fixed as 1 and the outer index determined from the profile in the envelope of the star at core collapse \citep{Takei_et_al_2022}. The velocity and density at the transition of the two power laws are determined from the total kinetic energy of the ejecta $E_{\rm ej}$ and the mass of the ejecta $M_{\rm ej}$. As our model is agnostic of the explosion mechanism, the value of $E_{\rm ej}$ is left for the user to specify when running \texttt{CHIPS}. The value of $M_\mathrm{ej}$ is calculated by subtracting the CSM mass $M_\mathrm{CSM}$ and the expected mass of the compact remnant $M_\mathrm{rem}$ from the progenitor mass,
\begin{eqnarray}
    M_\mathrm{ej}=M_*-M_\mathrm{CSM}-M_\mathrm{rem}(M_\mathrm{CO}),
\end{eqnarray}
where $M_\mathrm{rem}$ is a function of the mass of the carbon-oxygen (CO) core $M_\mathrm{CO}$ at core-collapse recorded in MESA. The helium star is typically formed by mass transfer in a close binary system \citep[e.g.,][]{Podsiadlowski_et_al_1992,Eldridge08,Yoon10}. We therefore use a formula of $M_\mathrm{rem}(M_{\rm CO})$ obtained for stripped stars \citep[][Case B mass transfer]{Schneider21}, \footnote{This formulation is ambiguous for our modeling of interacting Ic, since $x_{\rm CO}$ changes upon eruption. We however consider only progenitors that lead to neutron stars (NSs) according to this formula, where the NS mass is not sensitive to $x_{\rm CO}$ anyway.}
\begin{eqnarray}
&&\log_{10}\left(\frac{M_{\rm rem}}{M_\odot}\right) = \nonumber \\
&&\left\{ \begin{array}{ll}
\log_{10}(0.01909x_{\rm CO} + 1.34529) & (x_{\rm CO}< 7.548),\\
-0.02466x_{\rm CO}+1.28070 & (7.548 \leq x_{\rm CO}<8.491), \\
\log_{10}(0.01909x_{\rm CO} + 1.34529) & (8.491 \leq x_{\rm CO}<15.144),
\end{array}\right.
\label{eq:remnant_mass}
\end{eqnarray}
where $x_\mathrm{CO}$ is the mass of the CO core in units of solar mass. We plot $M_\mathrm{rem}$ as a function of $M_\mathrm{ZAMS}$ in the left panel of Figure \ref{fig:progenitors}, which takes almost the same value of $\approx1.4\,M_\odot$.

\subsubsection{Opacity in the LC code}
In order to solve the radiative transfer equations in the unshocked CSM, we need to calcluate the emissivity of gas in the unshocked CSM using the Planck mean opacities. The Planck mean opacities take values quite different from the Rosseland mean adopted in OPAL and \AE SOPUS. We therefore use the opacity tables for the Planck mean using TOPS \citep{1995ASPC...78...51M} as in the previous version of CHIPS, but extend it to hydrogen-free and helium-free gas. The temperature range for these tables is $3.7636\leq\log T\ {\rm [K]}\leq 6.0646$, while the range of $R$ is the same as the Rosseland mean tables. If $T$ and/or $R$ are outside the table, we use the values at the edge.

\subsubsection{Radioactive Decay of $^{56}\mathrm{Ni}$ and $^{56}\mathrm{Co}$}
\label{sec:56ni_co}
We also implement the thermal emission due to the radioactive decay of $^{56}\mathrm{Ni}$ and its daughter nucleus $^{56}\mathrm{Co}$.
The radiative flux from this emission is used as the flux at the immediate upstream of the reverse shock, which was set to 0 in the previous version of \cite{Takei_et_al_2022} that neglected $^{56}$Ni decay. Using this flux as the boundary condition at the reverse shock front, we follow the approach of \citet{Takei_et_al_2022} and integrate their Equations (12)--(14) to obtain the shock structure.
To obtain the time-dependent luminosity from radioactive heating, we follow the method described in Appendix A of \citet{Valenti_et_al_2008}. This method models photon diffusion in the ejecta for the initial photospheric phase, as well as the partial trapping of $\gamma$-rays and positrons in the later nebular phase\footnote{There is a typo in \citet{Valenti_et_al_2008} for the timescale $\tau_m$ (10/3 should be 6/5), which is also implied in the footnote in \cite{Wheeler15}.} \citep[for details, see also][]{Arnett82,Clocchiatti_Wheeler_1997}. The $^{56}\mathrm{Ni}$ mass $M_\mathrm{Ni}$ is specified by the argument \texttt{--Mni} in units of solar mass, and set to 0 by default. Although $M_{\rm Ni}$ in interacting SNe Ibn/Icn are inferred to be much smaller than those in SNe Ib/c \citep[e.g.,][]{Moriya_Maeda_2016}, it may still affect the light curve in the late phase. We note that this inclusion of $^{56}\mathrm{Ni}$ heating is approximate, and the assumptions on mixing of $^{56}\mathrm{Ni}$ and recombination in the ejecta may also affect the late-time light curves, and hence the inference of $M_{\rm Ni}$ \citep[e.g.,][]{Wheeler15,Haynie23}. 

\section{CHIPS Results}
\label{sec:Results}
Here we present some results of our calculation using the updated \texttt{CHIPS} code. We explore the dependences of the density profiles of the CSM, precursor light curves, and SN light curves on our key input parameters, i.e. ZAMS mass $M_\mathrm{ZAMS}$, the timing of eruption $t_\mathrm{inj}$, the injected energy $f_\mathrm{inj}$, and the nickel mass $M_\mathrm{Ni}$. To simplify our parameter space, we fix the explosion energy to $10^{51}\,{\rm erg}$ unless otherwise mentioned. We also compare our results with light curves of some observed precursors/SNe Ibn.

\subsection{Mass Eruption: Dependence on the Progenitor and Injected Energy}
\label{sec:mass_eruption}
\begin{figure*}
\centering
\includegraphics[width=\linewidth]{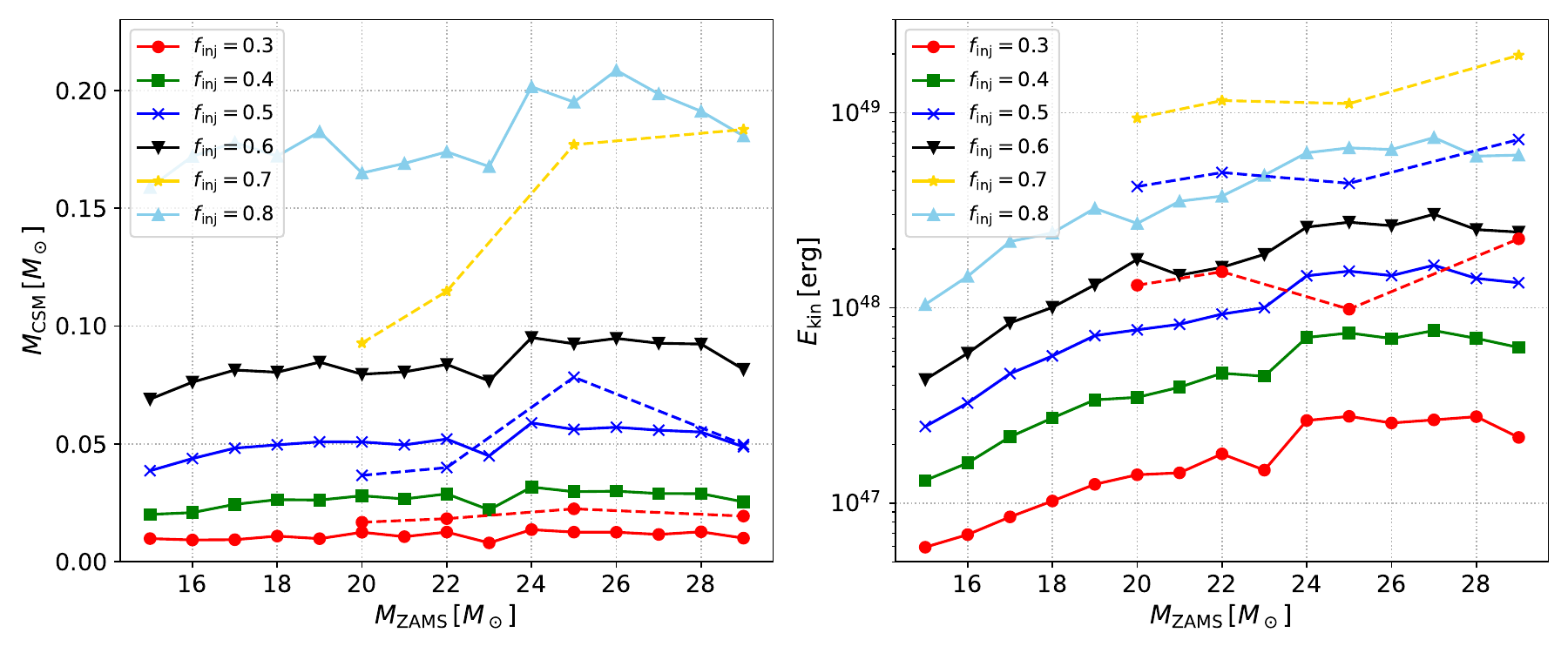}
\caption{The unbound CSM mass and the kinetic energy of the CSM as a function of $M_\mathrm{ZAMS}$ measured at $t=2t_\mathrm{dyn}$ for each $f_\mathrm{inj}$. This mass does not change much after $2t_\mathrm{dyn}$, because the thermal pressure no longer affects the expansion and fallback of the CSM. The corresponding values for helium-poor stars, which are possible progenitors of SNe Icn, are indicated by dashed lines.}
\label{fig:csmmass}
\end{figure*}
\begin{figure}
\centering
\includegraphics[width=\linewidth]{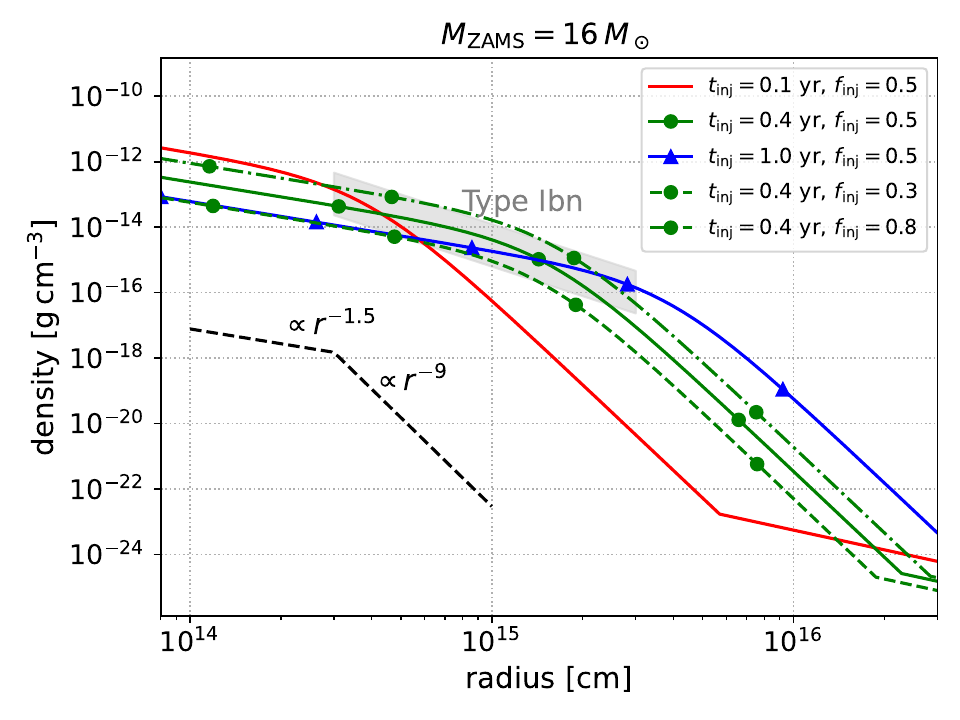}
\caption{Comparison of the density profile of the CSM just before the core-collapse as functions of the radius $r$. The selected parameters are $t_\mathrm{inj}=0.1,\,0.4,\,1\,{\rm yr}$, $f_\mathrm{inj}=0.3,\,0.5,\,0.8$, and a helium star with ZAMS mass of $16\,M_\odot$. The area shaded in gray represents the density profile inferred from light curve modeling of SNe Ibn \citep{Maeda_Moriya_2022}.}
\label{fig:csm_cmp}
\end{figure}

We plot the mass of the unbound CSM $M_\mathrm{CSM}$ as a function of $M_\mathrm{ZAMS}$ in Figure \ref{fig:csmmass} for each $f_\mathrm{inj}$. From this figure we clearly see the tendency that $M_\mathrm{CSM}$ increases with $f_\mathrm{inj}$, although the dependence on ZAMS mass is not trivial. The slightly higher $M_{\rm CSM}$ for stars of $M_{\rm ZAMS}=24$--$27\,M_\odot$ is probably caused by the larger mass of the helium layer for these progenitors (Figure \ref{fig:progenitors}).
We obtain the CSM mass of $\sim0.01-0.1\,M_\odot$ in the examined parameter space. The range of this value is consistent with previous estimation \citep{Maeda_Moriya_2022}.

We show the comparison of the density profiles of the CSM as functions of $r$ for different $t_\mathrm{inj},\,f_\mathrm{inj}$ in Figure \ref{fig:csm_cmp}. As can be clearly seen, the larger $f_\mathrm{inj}$, namely more energetic eruption, results in the formation of denser CSM. As for the dependence on $t_\mathrm{inj}$, the density of the CSM becomes low for larger $t_\mathrm{inj}$ because the CSM enters the homologous expansion phase immediately after mass eruption. As discussed in \citet{Tsuna21b,Takei_et_al_2022}, the density at a certain point is roughly proportional to $t_\mathrm{inj}^{-3}$. We also find that the density profiles plotted in the figure well agree with the profiles inferred from the observed SNe Ibn \citep{Maeda_Moriya_2022}.

\subsection{Mass Eruption: Precursor Emission}
\label{sec:precursor_result}
\begin{figure}
 \centering
 \includegraphics[width=\linewidth]{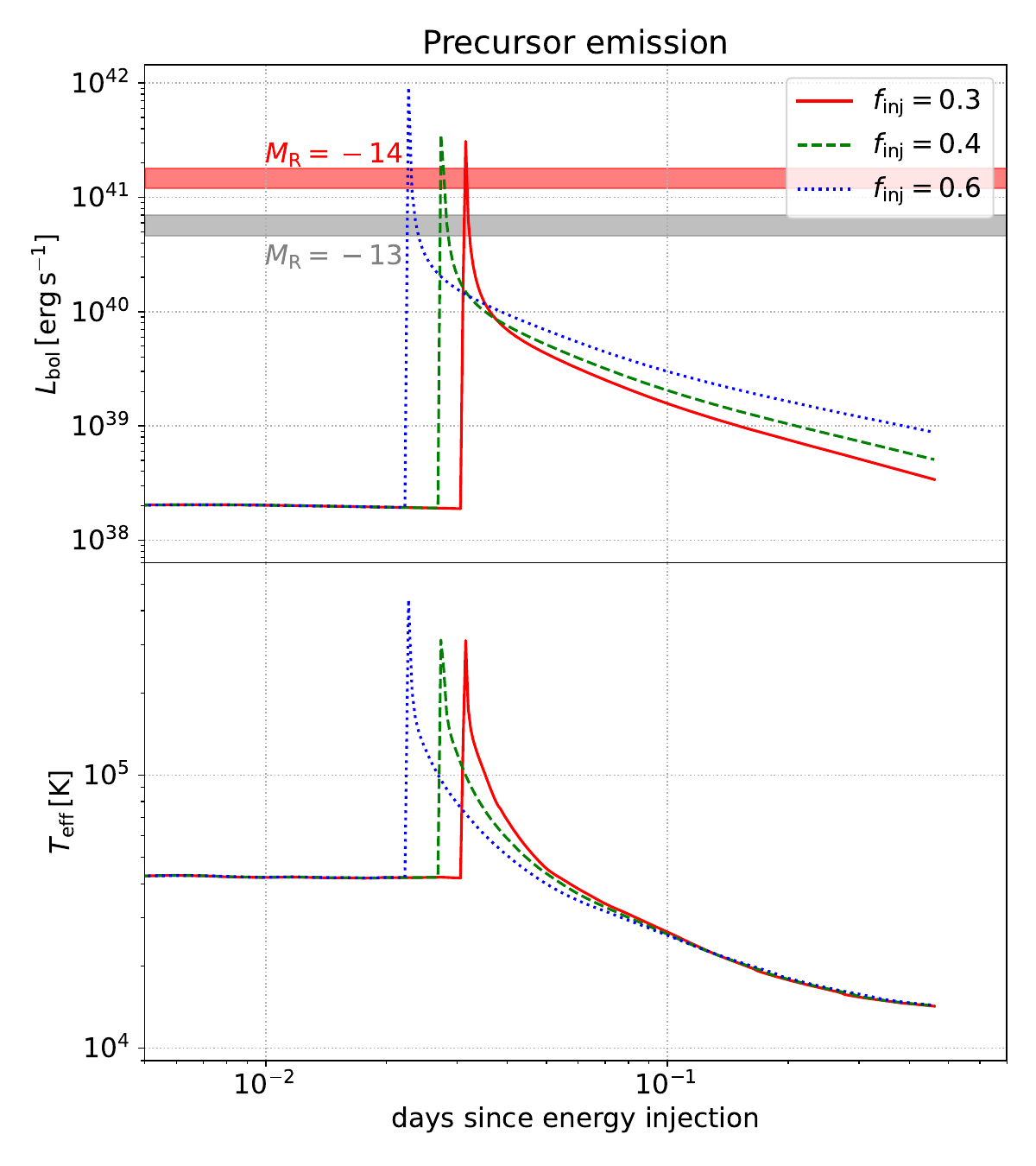}
 \caption{Top panel: Plots of the bolometric luminosity as a function of days since energy injection. The gray and red shaded regions represent the bolometric luminosity (assuming blackbody emission with temperature $5000$-$10000\,{\rm K}$) corresponding to the R band magnitude $M_\mathrm{R}=-13$ and $-14$ respectively, which are peak magnitudes of the precursor event prior to SN 2019uo \citep{Strotjohann_et_al_2021} and 2006jc \citep{Pastorello_et_al_2007}. Bottom panel: Plots of the effective temperature. We select a helium star of initial mass $M_\mathrm{ZAMS}=15\,M_\odot$, of which the radius is the largest in our progenitor models. }
 \label{fig:progenitor_emission}
\end{figure}

Figure \ref{fig:progenitor_emission} shows the light curves and temporal evolution of $T_\mathrm{eff}$ for a $15\,M_\odot$ helium-star model with $f_\mathrm{inj}=0.3,\,0.4,\,0.6$. The larger $f_\mathrm{inj}$ results in the more luminous precursor.  The luminosity quickly reaches the peak at typically only $10^{39}$--$10^{40}$ erg s$^{-1}$ and declines to a brightness comparable to that of the original progenitor within a day. 

To our knowledge, the precursor events for hydrogen-poor SNe are observed only for five events including two Type I superluminous SNe \citep[SN 2006jc, LSQ14bdq, SN 2018gep, SN 2019uo, SN 2019szu;][respectively, see also \citealt{Corsi14} for a candidate event]{Pastorello_et_al_2007, Nicholl_et_al_2015, Ho19, Strotjohann_et_al_2021, Aamer_et_al_2023}, with peak brightness from $-13$ to $-14$ mag in the optical declining for (1--few) $\times$ 10 days to the detection limit. The light curve duration of mass eruption we see here is orders of magnitude shorter than the observed precursors, and the luminosity after the brief peak is also an order of magnitude fainter. As discussed in \citet{Kuriyama20a}, the duration of the light curve is mainly determined by the progenitor radius. In the case of compact stars, a large fraction of the injected thermal energy is consumed to expand the layer instead of being radiated away. Therefore, we need an alternative model that generates internal energy at a larger radius to explain the observed long-lasting precursor. 

Collision of the ``precursor ejecta" with an external material is one natural explanation, as discussed intensively throughout this paper. The external CSM can either be a compact one or an extended one, depending on whether the extension is smaller or larger than the diffusion radius $R_d\sim \kappa \rho r^2 v_{\rm sh}/c$ according to \cite{Chevalier_Irwin_2011}. In the former case the precursor would be powered by shock cooling emission (i.e. radiation of optically thick shocked CSM as it expands and cools), while in the latter case it would be powered by prolonged CSM interaction (i.e. gradual conversion of the kinetic energy in the precursor ejecta into radiation).
Given that the precursor occurred at $\sim$ year before explosion, the latter case of extended CSM may be more natural. For the case of compact CSM, the interval of two successive enhanced mass loss has to be much shorter than this, which is not impossible but needs fine-tuning.

We analytically estimate the precursor luminosity\footnote{Simulations of two successive eruptions with CHIPS have been carried out for supergiants \citep{Kuriyama21,Tsuna_et_al_2023}, but for the case here the interval between eruptions being orders of magnitude longer than $t_{\rm dyn}$ precludes such a simulation. We furthermore attempted to calculate using the light curve part of CHIPS, but the low shock velocities resulted in numerical difficulties at the early phase of the interaction.}, based on the properties of the erupted material in Section \ref{sec:mass_eruption}. For typical Ibn progenitors, the velocity of the bulk of the ejecta is estimated from Figure \ref{fig:csmmass} to be $\sim$ a few $1000$ km s$^{-1}$. Assuming this crashes to a pre-existing material with velocity $v_{\rm ext}$ and density $\rho_{\rm ext}$, the instantaneous luminosity by shock dissipation is
\begin{eqnarray}
        L_{\rm sh}&=& \epsilon \times 2\pi r_{\rm sh}^2 \rho_{\rm ext} (v_{\rm sh}-v_{\rm ext})^3 \nonumber \\
        &\approx& 5\times 10^{40}\ {\rm erg\ s^{-1}}\left(\frac{\epsilon}{0.25}\right) \left(\frac{r_{\rm sh}}{2\times 10^{14}\ {\rm cm}}\right)^2 \nonumber \\
        &&\times \left(\frac{\rho_{\rm ext}}{10^{-13}\ {\rm g\ cm^{-3}}}\right) \left(\frac{v_{\rm sh}-v_{\rm ext}}{2000\ {\rm km\ s^{-1}}}\right)^3,
\end{eqnarray}
where we adopt $\epsilon\approx 0.25$ as a typical conversion efficiency into radiation \citep{Takei20}. The much lower velocity of the shock than normal SNe realizes an efficient conversion of the dissipated shock energy to photons in the optical band \citep{Svirski12,Chevalier12,Tsuna20}. The luminosity is roughly consistent with the precursors observed in the Type Ibn SN 2006jc and 2019uo with luminosity of $\sim (5-10) \times 10^{40}$ erg s$^{-1}$, if the erupted material collides with an extended material of $\rho_{\rm ext}\sim 10^{-13}\ {\rm \ cm^{-3}}$ at a few $\times 10^{14}$ cm. If we assume a wind profile $\rho_{\rm ext}=\dot{M}_{\rm ext}/4\pi r^2 v_{\rm ext}$, this corresponds to a mass-loss rate of
\begin{eqnarray}
    \dot{M}_{\rm ext}&\sim& 0.08 M_{\odot}\ {\rm yr}^{-1} \nonumber \\
    &\times &\left(\frac{r}{2\times 10^{14}\ {\rm cm}}\right)^2 
    \left(\frac{\rho_{\rm ext}}{10^{-13}\ {\rm g\ cm^{-3}}}\right)  \left(\frac{v_{\rm ext}}{1000\ {\rm km\ s^{-1}}}\right).
\end{eqnarray}

The precursor should fade either when the precursor shock exits the external material, or when the shock propagates to a radius where $v_{\rm ext}$ becomes comparable to the shock velocity if $v_{\rm ext}$ increases with radius like the CSM from mass eruption. The Ibn precursors last for 10--30 days, which implies a CSM extent of at least $(2-6)\times 10^{14}$ cm $[(v_{\rm sh}-v_{\rm ext})/2000$ km s$^{-1}]$ at the onset of the precursor. 
While the origin of the external material is unclear, according to Figure \ref{fig:csm_cmp} such values for (the averaged) $\dot{M}_{\rm ext}$ and radius are reproduced if another eruption had happened $\lesssim$ 0.5 yr before the precursor event.

We did not consider the attenuation by dust that may be formed by the enhanced mass loss prior to the precursor event. For a relatively dim precursor a significant amount of dust can survive from sublimation by the precursor emission \citep[e.g.,][]{Neustadt23}, which may explain the non-detection of precursors in the optical for the recent SN 2023ixf had there been one \citep{Dong23,Hiramatsu23}. 

To see if dust may sublimate by the precursor, we estimate the sublimation radius $R_\mathrm{c}$. This radius can be obtained from the optical/UV luminosity $L_\mathrm{opt/UV}$ and the critical temperature $T_\mathrm{c}$ above which dust evaporates by radiation \citep{Dwek1985,Omand19},
\begin{eqnarray}
    R_\mathrm{c}&=&\left(\frac{L_\mathrm{opt/UV}}{16\pi \sigma_\mathrm{SB}T_\mathrm{c}^{4}}\frac{Q_\mathrm{opt/UV}}{\langle Q\rangle_{T_\mathrm{c}}}\right)^{1/2} \nonumber \\
    &\approx& 2.3\times 10^{15}\,{\rm cm}\left(\frac{L_\mathrm{opt/UV}}{10^{41}\,{\rm erg\,s^{-1}}}\right)^{1/2} \left(\frac{T_\mathrm{c}}{2300\,{\rm K}}\right)^{-2} \nonumber \\
        &&\times \left(\frac{Q_\mathrm{opt/UV}/1}{\langle Q\rangle_{T_\mathrm{c}}/0.23}\right)^{1/2},
\end{eqnarray}
where $\sigma_\mathrm{SB},\,Q_\mathrm{opt/UV},\,\langle Q\rangle_{T_\mathrm{c}}$ denote the Stefan-Boltzmann constant, the absorption efficiency factors averaged over the optical/UV spectrum, and that averaged over all frequency, respectively.
Thus if the dense matter existing before the precursor is confined within $\sim 10^{15}$ cm, the dust that may exist there would be sublimated and would not much affect the luminosity estimated above. Dust located beyond this radius can survive and absorb the optical/UV emission from the precursor and re-emit in near-infrared bands. 
In fact, evidence for pre-existing dust at a much larger radius of $\gtrsim3\times 10^{16}\,{\rm cm}$ has been found from observations of one SN Ibn OGLE-2012-SN-006 \citep{Gan_et_al_2021}.
 
\subsection{Supernova Light Curves}
In this subsection, we present the dependence of the characteristic observables of the SN light curves, such as the peak luminosity $L_\mathrm{peak}$, the rise time $t_\mathrm{rise}$ on parameters $t_\mathrm{inj},\,f_\mathrm{inj},\,M_\mathrm{Ni}$. Here we limit our discussion to the helium star progenitors, but the conclusion is similar for the helium-poor progenitors.

\subsubsection{Dependence on the Timing of Eruption $t_\mathrm{inj}$}
\begin{figure}
\centering
\includegraphics[width=\linewidth]{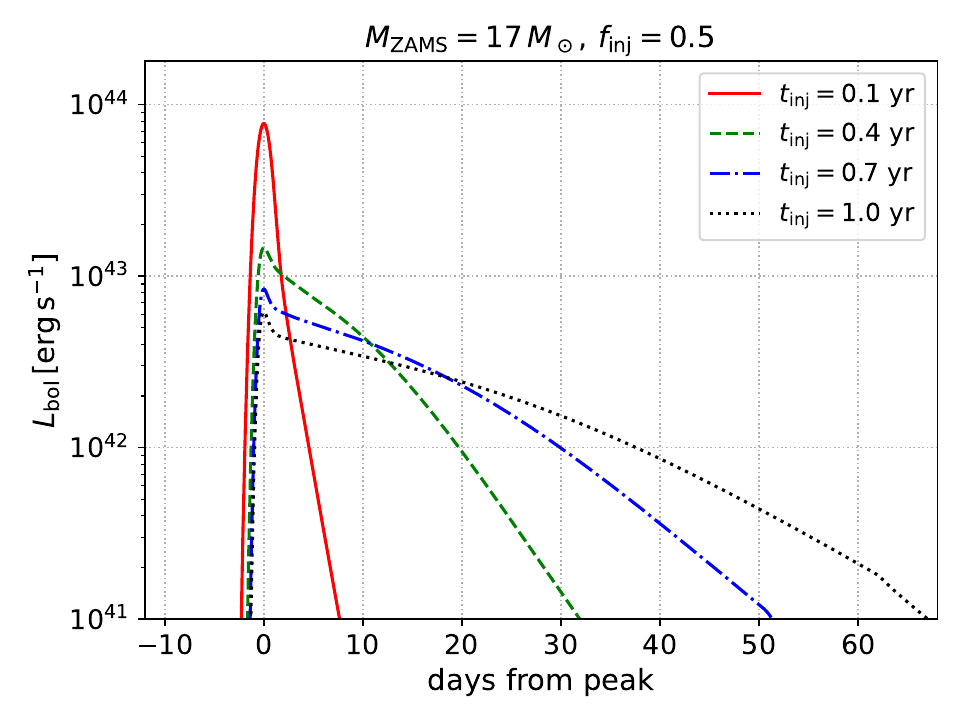}
\caption{Sample bolometric light curves for the $f_{\rm inj}=0.5,\,M_\mathrm{ZAMS}=17\,M_\odot$ helium star model with varied $t_\mathrm{inj}$. $M_\mathrm{Ni}$ is fixed to $0$.}
\label{fig:lc_cmp_dep_tinj}
\end{figure}
In Figure \ref{fig:lc_cmp_dep_tinj} we plot the bolometric light curves for $t_\mathrm{inj}=0.1,\,0.4,\,0.7,\,1\,{\rm yr}$. While $t_\mathrm{rise}$ does not largely depend on $t_\mathrm{inj}$, $L_\mathrm{peak}$ increases with $t_\mathrm{inj}$. The rise time depends on the shock breakout radius $R_d$ and the shock velocity $v_\mathrm{sh}$ \citep{Chevalier_Irwin_2011}. Although larger $t_\mathrm{rise}$ is obtained in our previous work for SNe IIn, we cannot reproduce such a longer rise time in the parameter spaces we explore. This is mainly caused by the low scattering opacity due to the high ionization energy of helium and small CSM masses. In contrast, we obtain $t_\mathrm{rise}$ of $\sim4\,{\rm day}$ for models of SNe Icn, since free electrons from carbon and oxygen contribute to the larger scattering opacity.
 
Since smaller $t_\mathrm{inj}$ results in denser CSM and smaller radius, the peak luminosity and decline rate become larger.

\subsubsection{Dependence on the Injecting Energy $f_\mathrm{inj}$}
\begin{figure}
\centering
\includegraphics[width=\linewidth]{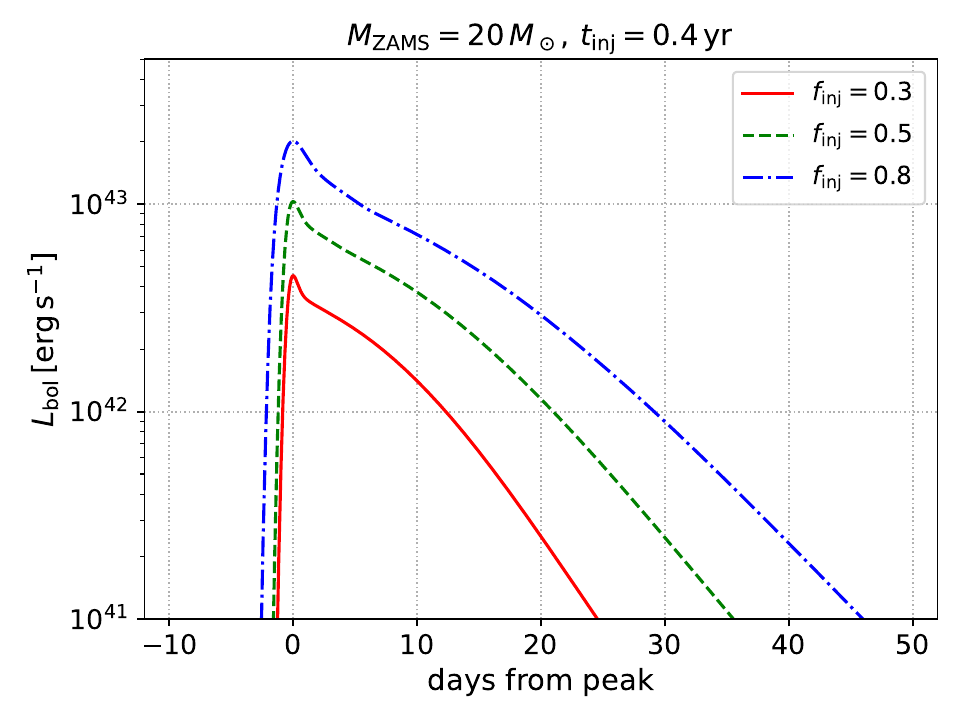}
\caption{Sample bolometric light curves for the $t_{\rm inj}=0.4\,{\rm yr},\,M_\mathrm{ZAMS}=20\,M_\odot$ helium star model with varied $f_\mathrm{inj}$. $M_\mathrm{Ni}$ is fixed to $0$.}
\label{fig:lc_cmp_dep_finj}
\end{figure}
Figure \ref{fig:lc_cmp_dep_finj} shows the dependence of light curves on $f_\mathrm{inj}$. As mentioned in Section \ref{sec:mass_eruption}, higher $f_\mathrm{inj}$ results in the larger mass eruption. Since the radiated energy of interaction-powered SNe increases with $M_\mathrm{CSM}$ \citep[e.g.,][]{van_Marle_et_al_2010,Moriya_et_al_2011}, the peak luminosity is higher for larger $f_\mathrm{inj}$.
The duration is also longer for larger $f_\mathrm{inj}$, because the velocity of the CSM is higher and thus the CSM is more extended to the larger radius for given $t_\mathrm{inj}$.

\subsubsection{Dependence on the Nickel Mass $M_\mathrm{Ni}$}
\begin{figure}
\centering
\includegraphics[width=\linewidth]{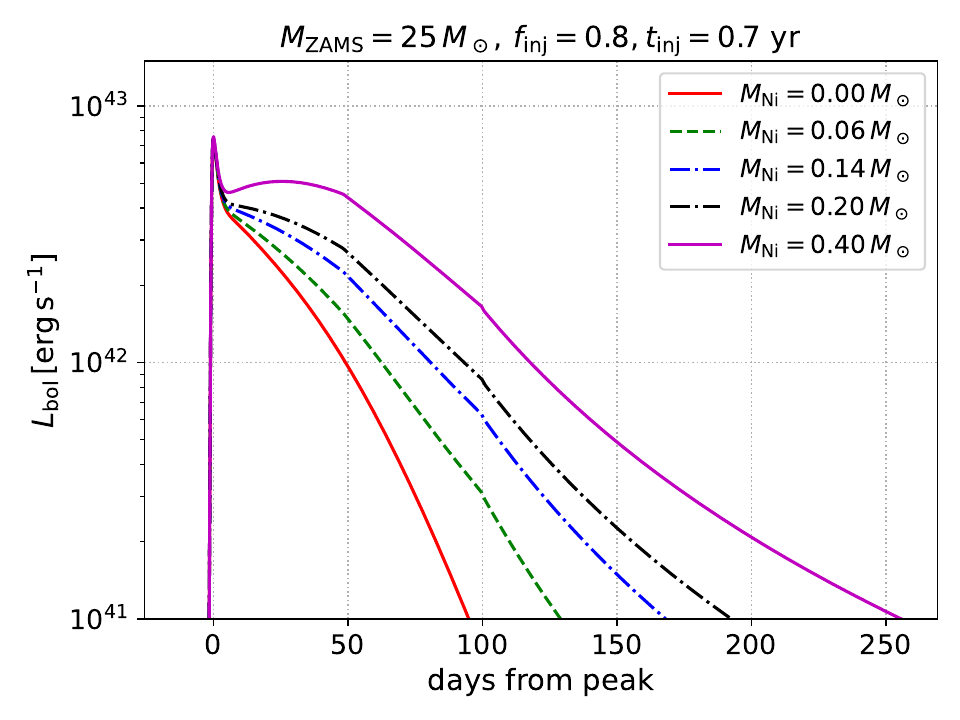}
\caption{Bolometric light curves for the $M_\mathrm{ZAMS}=25\,M_\odot$ helium star model with $f_{\rm inj}=0.8, t_{\rm inj}=0.7\,{\rm yr}$ model with varied mass of $^{56}$Ni. The explosion energy of the SN is fixed to $10^{51}$ erg.}
\label{fig:ni56}
\end{figure}

We plot the dependence of light curves on the nickel mass $M_\mathrm{Ni}$ in Figure \ref{fig:ni56}. Due to the high conversion efficiency, the peak luminosity is mainly determined by the interaction. On the other hand, the thermal emission from the radioactive decay of $^{56}$Ni can be dominant at the late phase, because the contribution from CSM interaction generally decays rapidly. The shape of light curves depends also on $t_\mathrm{inj}$, as this controls the duration of the CSM interaction. 

Bumpy structures in the light curve can emerge, depending on $M_\mathrm{Ni}$ or timescale of the interaction. For a large nickel mass, the light curve can even display a separate peak powered by the radioactive decay that follows the first peak due to CSM interaction. We find that a second peak is typically seen for $M_{\rm Ni} \gtrsim 0.4\,M_\odot$, which is within the range of values inferred from
broad-lined Type Ic SNe \citep[e.g.,][]{Drout11,Prentice16,Taddia18,Taddia19}. As explained in the next section, such double peaked SN Ibc are starting to be discovered by high-cadence surveys \citep{Kuncarayakti_2023,Das23}.

\subsubsection{Comparison with Observations}
\label{sec:observation}
\begin{figure*}
\centering
\includegraphics[width=\linewidth]{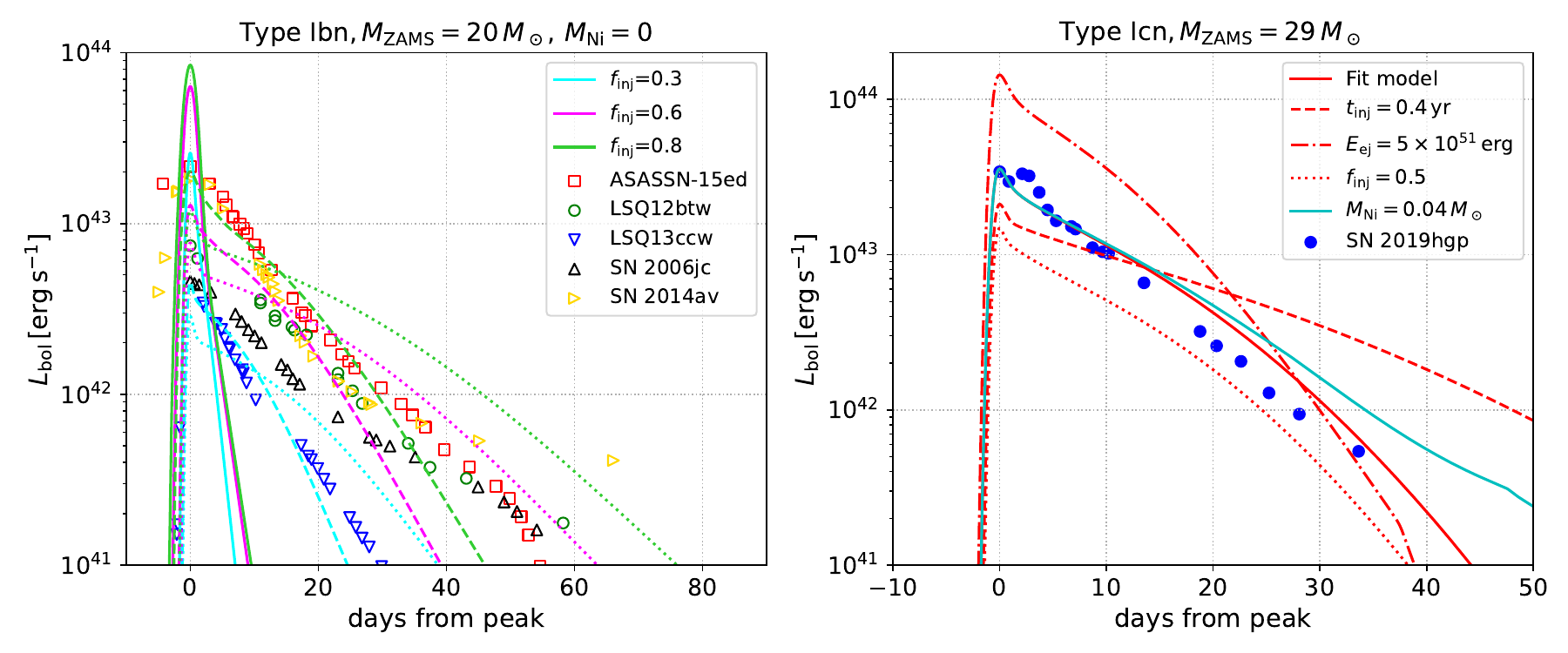}
\caption{Left panel: Comparison of our results with compilation of observed SNe Ibn. The solid lines represent models with $t_\mathrm{inj}=0.1\,{\rm yr}$, dashed lines with $t_\mathrm{inj}=0.4\,{\rm yr}$, and dotted lines with $t_\mathrm{inj}=0.7\,{\rm yr}$. $M_\mathrm{Ni}$ is fixed to 0. Right panel: A fit model for SN Icn 2019hgp. The selected parameters are $E_\mathrm{ej}=2\times10^{51}\,{\rm erg},\,f_\mathrm{inj}=0.7,\,t_\mathrm{inj}=0.23\,{\rm yr}, M_\mathrm{Ni}=0$. For comparison, we also plot models with only one parameter changed out of the four. The changed parameter is shown in the legend.}
\label{fig:cmp_w_obs_sn}
\end{figure*}
Here we compare our \texttt{CHIPS} results with observational data of SNe Ibn 2006jc \citep{Pastorello_et_al_2007}, LSQ12btw, LSQ13ccw \citep{Pastorello_et_al_2015_ccw_btw}, 2014av \citep{Pastorello_2016_2014av}, ASASSN-15ed \citep{Pastorello_2015_ASASSN_15ed}, and SN Icn 2019hgp \citep{Gal-Yam_et_al_2022}.

In Figure \ref{fig:cmp_w_obs_sn}, we plot bolometric light curves varying $f_\mathrm{inj},\,t_\mathrm{inj}$ together with the observations of interacting SNe Ibc.
As can be seen from the left panel, the light curve features of SNe Ibn, such as the rise time, peak luminosity, and decline rate, are generally reproduced by \texttt{CHIPS}. The typical rise time of our light curve models is one to a few days, which is in agreement with LSQ13ccw with a well-sampled peak, as well as Ibn events from recent high-cadence surveys that capture the light curve rise \citep{Ho23}.

However, some of the events like SN 2014av display a longer rise time of $\approx 10$ days, which we struggle to reproduce with our light curve model. Another slowly-evolving SN Ib with signatures of interaction is OGLE-2012-SN-006 \citep{Pastorello_et_al_2015_OGLE}, which has a rise time of $\sim10\,{\rm days}$ in the $I$ band. The decline rate in the late phase matches the expectation from $^{56}\mathrm{Co}$ decay, although this match is coincident and the flattening of the light curve after peak in $I$ band is likely caused by dust formation, indicating that extinction in optical/UV bands also occurs at early phase \citep{Gan_et_al_2021}. We speculate that such pre-existing dust, with its large optical depth, may play a role in prolonging the light curve rise in the optical. We note that dust can cause substantial reddening in the rise, and may be disfavored for some (e.g. SN 2010al, ASASSN-15ed) that have rather blue color at early times. Alternatively, the slow rise can be realized by a CSM even more massive than the range probed in our calculations. Detailed multi-band light curve modeling with dust physics would be important to discriminate these possibilities.

We next investigate the light curves of Type Icn SNe using our helium-free progenitors, with SN 2019hgp \citep{Gal-Yam_et_al_2022} as an example. We show the comparison of our result for parameters $M_\mathrm{ZAMS}=29\,M_\odot,\,E_\mathrm{ej}=2\times10^{51}\,{\rm erg},\,f_\mathrm{inj}=0.7,\,t_\mathrm{inj}=0.23\,{\rm yr},\,M_\mathrm{Ni}=0$ with SN 2019hgp in the right panel of Figure \ref{fig:cmp_w_obs_sn}. The CSM mass of this model, $\approx0.18\,M_\odot$, is roughly consistent with the simple estimation by \citet{Gal-Yam_et_al_2022}.

For the samples compared here, our models require only small amounts of $^{56}$Ni, if any, to explain the late phase of the light curve.  This is in line with previous studies on SNe Ibn and Icn \citep{Moriya_Maeda_2016,Maeda_Moriya_2022,Perley22,Pellegrino22}, which found $M_{\rm Ni}$ to be much smaller than stripped-envelope SNe without interaction signatures. However, there is a growing population of Type Ibc SNe whose optical light curves appear as double-peaked, which may be powered by both $^{56}$Ni and CSM interaction \citep[e.g.,][]{Nakar_Piro_14,Kuncarayakti_2023,Das23}. Our code is suited for numerical modeling of such transients, and we plan to explore such possibilities in detail in future work.

\section{Summary and Discussion}
\label{sec:conclusion}
In this work we have updated our \texttt{CHIPS} code, which can model SN powered by CSM interaction as well as the mass outburst that creates the CSM, with the aim of modeling interacting hydrogen-poor SNe.
The parameters that the user can tune are the progenitor model, the time from mass eruption to core-collapse $t_\mathrm{inj}$, the injected energy scaled with the envelope's binding energy $f_\mathrm{inj}$, the explosion energy $E_\mathrm{ej}$, and the nickel mass synthesized at the time of core-collapse $M_\mathrm{Ni}$, which is newly implemented in this update.
We find that the explosion of a stripped progenitor embedded in a CSM of mass $0.01$--$0.1\,M_\odot$ ejected $t_\mathrm{inj}=0.1$--$0.7\,{\rm yr}$ before explosion can generally explain the features of SNe Ibn/Icn light curves, such as the peak luminosity and decline rate. For the rise time, a few SNe Ibn events show a much slower rise than our model predictions.
Besides, the density profile of the CSM formed by mass eruption is consistent with that obtained from independent modeling of SN Ibn/Icn \citep{Maeda_Moriya_2022}.
We also find that the observed precursors of SN Ibn can be explained by the outburst that creates the dense CSM, given that the energy of the outburst is efficiently dissipated by collision with an external material, possibly generated by a previous mass eruption.

In what follows, we conclude by briefly discussing the possible mechanisms for enhanced mass loss, and some caveats of our model.

\subsection{Implications for Mass Loss Scenarios}
One of the plausible mechanisms that can explain the dense CSM is the pulsational pair instability (PPI) caused by the production of electron-positron pairs \citep{Heger02,Yoshida16,Woosley_2017,Woosley2019,Leung19,Renzo_et_al_2020}.
If pair production occurs at the center of a star, such an event can create CSM mass of $>0.1\,M_\odot$, depending on the helium core mass. \citet{Woosley_2017} suggests that PPISNe are possible candidates for interaction-powered transients such as Type IIn, Ibn/Icn.  
This scenario can have multiple eruptions in a short timescale of years or less, which can be a natural explanation for the events with precursors as we discussed in Section \ref{sec:precursor_result}. A potential difficulty in this scenario for explaining the entirety of SN Ibn is that the light curve of SN Ibn events are quite similar in morphology \citep{Hosseinzadeh_et_al_2017}, in contrast to SN IIn. This may be at odds with the orders of magnitude diversity in mass and time interval of each ejection expected from these models. However, a more detailed light curve modeling would be needed to test this possibility.

Another possible mechanism is the off-center silicon flash which occurs in helium stars with initial masses of 2.5--3.2\,$M_\odot$ \citep{Woosley2019}. The mass of the ejected envelope that becomes CSM is found as $M_{\rm CSM}=0.01-0.73\,M_\odot$, mainly depending on how much silicon burns (which is linked to the injected thermal energy in our model). Due to the flash lagging the final core-collapse, the terminal explosion occurs $t_\mathrm{inj}=0.04$--$0.2$\,yr after the mass eruption, with $t_{\rm inj}$ being larger for larger $M_{\rm CSM}$. This parameter space apparently do not overlap with what is required to reproduce the typical fast-rising SN Ibn, as the ones with largest $t_{\rm inj} \gtrsim 0.1$ yr tend to have ejected masses much larger than $0.1\,M_\odot$.
Nevertheless this channel has a potential to explain SN Ibn with long rise times, which we struggle to reproduce in this study. The explosion energies of these progenitors are expected to be lower than the canonical $10^{51}$ erg, which may also aid to explain the slow evolution. We could not explore the scenario in detail, as helium stars in this mass range were difficult to generate by our \texttt{MESA} simulations due to the severe limitation of the time step at around the neon/oxygen burning phase.

Apart from the above single star cases, binary interactions may also play an important role in forming the dense CSM. A part of the orbital energy can be released to invoke mass ejection if the stripped star is in a close binary system with a compact object \citep{Chevalier_2012,Metzger_2022,Wu22_Ibc}. Such interaction may be timed with the explosion if an explosion is triggered by the compact object spiraling into the stellar core \citep[e.g.,][]{Fryer98,Soker19,Schroder20}. Following the discussion in \citet{Ko_et_al_2022}, we estimate the energy injection rate by the binary interaction. A close binary system consisting of a massive star with CO core mass of $M_\mathrm{CO}$ and a compact star with mass of $M_\mathrm{rem}$ which spirals in at the base of the helium layer has the orbital energy $E_\mathrm{orb}$ and period $t_\mathrm{Kep}$,
\begin{eqnarray}
    E_\mathrm{orb}&=&G\frac{M_\mathrm{CO}M_\mathrm{rem}}{2r_\mathrm{in}} \nonumber \\
    &\approx& 5.7\times 10^{49}\,{\rm erg}\left(\frac{M_\mathrm{CO}}{3\,M_\odot}\right) \left(\frac{M_\mathrm{rem}}{1\,M_\odot}\right)\left(\frac{r_\mathrm{in}}{0.1\,R_\odot}\right)^{-1},\\
    t_\mathrm{Kep}&=&2\pi\left[\frac{r_\mathrm{in}^{3}}{G(M_\mathrm{CO}+M_\mathrm{rem})}\right]^{1/2} \nonumber \\
    &\approx& 160\,{\rm s}\left(\frac{M_\mathrm{CO}+M_\mathrm{rem}}{4\,M_\odot}\right)^{-1/2} \left(\frac{r_\mathrm{in}}{0.1\,R_\odot}\right)^{3/2},
\end{eqnarray}
where $r_\mathrm{in}$ denotes the radius of the boundary between the CO core and helium layer.
For instance, our progenitor model with $M_\mathrm{ZAMS}=20\,M_\odot$ has an envelope binding energy of $\approx (-3.0\times10^{49})\,{\rm erg}$ and $r_\mathrm{in}\approx 0.12\,R_\odot$.
The fraction of the binding energy used to expel the envelope is commonly parameterized as the $\alpha$ parameter. These order-of-magnitude estimates indicate that $\alpha\gtrsim 0.1$ would correspond to $f_{\rm inj}\gtrsim 0.2$, and is sufficient for significant mass ejection.

The calculations done in this paper qualitatively differ from calculations usually done in the context of wave heating for stripped stars \citep{Fuller_Ro_2018,Leung_et_al_2021,Wu_Fuller_2022}. In this scenario, the energy carried as gravity waves are predicted to generally dissipate near the surface, in contrast to energy injection near the core that was considered in this paper. We note that our code is flexible in setting the location of the energy injection, so modeling mass loss mimicking this situation is in principle possible. Meanwhile, the more recent modelings \citep{Leung_et_al_2021,Wu_Fuller_2022} found that wave heating alone is energetically difficult to create a large mass of CSM required to explain SN Ibn, and additional processes, such as subsequent binary interactions, may be needed to drive intense mass-loss.

\subsection{Possible Caveats}
Our code is entirely one-dimensional, which neglects potential multi-dimensional effects in the generation of the CSM as well as the interaction itself. Polarimetry of Type Ibn SN 2015G showed high asymmetry in the CSM \citep{Shivvers17}, while similar observations for Type Icn SN 2021csp \citep{Perley22} and Type Ibn SN 2023emq \citep{Pursiainen_2023} found morphologies close to spherical symmetry. While measurements of asymmetry in the CSM in Type Ibn/Icn SNe are still limited, future surveys would increase this number to a great extent. This may become a strong clue for the origin of the CSM, and we plan to develop and carry out multi-dimensional calculations in the future.

Another noteworthy caveat in our code is that we do not take into account the effect of dust formed before/after the SN explosion.
Some SNe Ibn are known to have newly formed and/or preexisting dust from IR excess detection \citep{Smith_et_al_2008,Mattila_et_al_2008,Gan_et_al_2021}. As discussed in Section \ref{sec:precursor_result}, dust attenuates the optical emission and re-emits IR radiation. Taking this effect into account, the observed bolometric luminosity can be larger because it is generally constructed only from the optical and near-infrared $UBVRI$ bands, which means that our models may overestimate the bolometric luminosity compared to observationally derived one.

It is known that candidates for the location of dust are the unshocked ejecta/CSM, and the cool dense shell (CDS) formed in the shocked region. For dust formation in the unshocked ejecta, heating by energetic particles from the radioactive elements such as $^{56}\mathrm{Ni},\,^{56}\mathrm{Co},\,^{44}\mathrm{Ti}$ may be important as it inhibits dust condensation \citep{Sarangi_2013}. 
Another realistic situation is dust formation in CDS. From observations of SN 2006jc, the attenuation by dust $\sim50\,{\rm days}$ after its explosion is likely to be caused by the newly formed dust in the CDS \citep{Smith_et_al_2008}. The interplay of various physics involving dust formation/destruction in the ejecta is still quite uncertain \citep[][and references therein]{Kirchschlager19}, and simulating the CDS require resolving the shocked region with two-temperature radiation hydrodynamics simulations.

Finally, our code assumes Newtonian dynamics and does not treat relativistic fluid motion.
Since the progenitors of SNe Ibn/Icn are considered to be compact stars, the outer edge of the SN ejecta might be relativistic \citep[e.g.,][]{Matzner99,Tan_et_al_2001}. 
However, for a typical explosion energy of $E_\mathrm{ej}=10^{51}\,{\rm erg}$ that reproduces Type Ibn/Icn, the relativistic part of the ejecta ($\Gamma\beta>1$, where $\beta=v/c$ with $c$ being the speed of light and $\Gamma=1/\sqrt{1-\beta^2}$ is the Lorentz factor) is expected to have a kinetic energy of only $\ll 10^{47}\,{\rm erg}$. As this is much smaller than the total shock-dissipated energy of $\sim10^{49}$--$10^{50}\,{\rm erg}$ around the light curve peak, neglecting relativistic effects can be justified for our modeling of optical/UV emission from canonical ($E_{\rm ej}\sim 10^{51}$ erg) explosions of compact stars. It is noted that relativistic effects may be important for modelling photons of higher energies. For example, the collision of the relativistic part of the ejecta with a dense CSM may result in a low-luminosity gamma-ray burst \citep[e.g.,][]{Nakar_Sari_2012,Nakar15,Suzuki_et_al_2019} that mainly emit in X-rays and gamma-rays.

\begin{acknowledgements}
The authors thank the anonymous referee for valuable comments that greatly improve the manuscript.
This work is supported by JSPS KAKENHI Grant Numbers 21J13957, 22K03688, 22K03671, 20H05639, MEXT, Japan. YT is profoundly grateful to Haruki Shono and Kazuki Teraoka for financial supports. DT is supported by the Sherman Fairchild Postdoctoral Fellowship at Caltech.
TK is supported by the RIKEN Junior Research Associate Program. 
\end{acknowledgements}

\software{MESA \citep[v12778;][]{Paxton11,Paxton13,Paxton15,Paxton18,Paxton19,Jermyn23}, CHIPS \citep[v1.0.1;][]{Takei_et_al_2022}, Python libraries: Matplotlib \citep[v3.4.2;][]{Hunter:2007}, Numpy \citep[v1.17.4;][]{harris2020array}, Scipy \citep[v1.11.1;][]{2020SciPy-NMeth}}

\appendix
\section{Dependence on Duration of Energy Injection}
\label{sec:dependence_duration}
\begin{figure*}
\centering
\includegraphics[width=\linewidth]{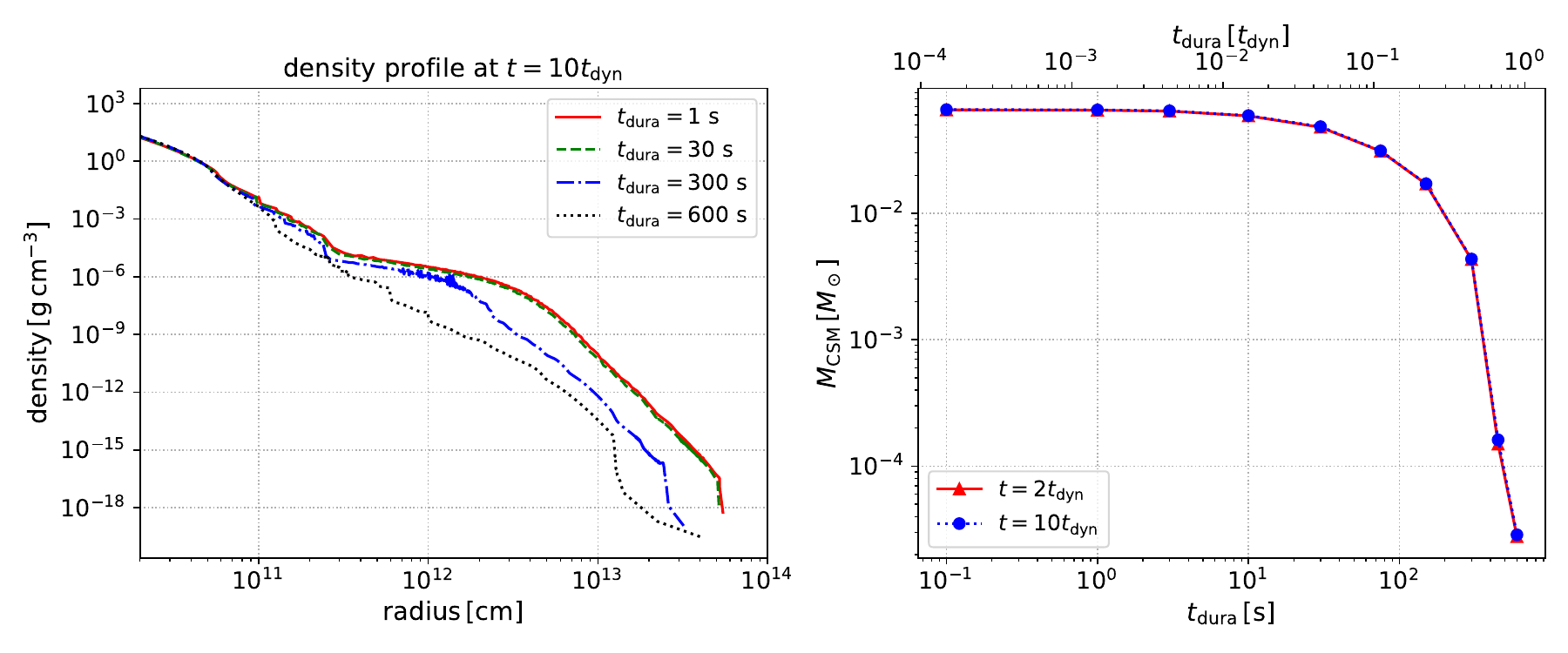}
\caption{Left: Comparison of the density profiles at $t=2t_\mathrm{dyn}$ for different $t_\mathrm{dura}$. Right: The unbound CSM mass as a function of $t_\mathrm{dura}$ at different epoch $t=2t_\mathrm{dyn},\,10t_\mathrm{dyn}$.}
\label{fig:csmmass_vs_tdura}
\end{figure*}
In this Appendix, we examine the response of the helium envelope to the continuous energy injection. This is important because the duration of the energy injection depends on its mechanism \citep[e.g.,][]{Woosley2019,Renzo_et_al_2020,Leung_et_al_2021}.
We adopt the progenitor model with mass of $M_\mathrm{ZAMS}=24\,M_\odot$ ($t_\mathrm{dyn}\approx676\,{\rm s}$), and inject the energy of $f_\mathrm{inj}=0.5$ over a period of $t_\mathrm{dura}=0.1,\,1,\,10,\,30,\,75,\,150,\,300,\,450,\,600\,{\rm s}$.

In Figure \ref{fig:csmmass_vs_tdura} we plot the density profile of models with $t_\mathrm{dura}=1,\,30,\,300,\,600\,{\rm s}$ and $M_\mathrm{CSM}$ as a function of $t_\mathrm{dura}$. As can be seen from the figure, $t_\mathrm{dura}$ much shorter than $t_\mathrm{dyn}$ results in almost the same density profile and thus the same CSM mass of $\sim0.07\,M_\odot$. In contrast, the CSM mass rapidly drops when $t_\mathrm{dura}$ is comparable to $t_\mathrm{dyn}$, which is consistent with a previous work \citep{Ko_et_al_2022}.

We have also confirmed that the unbound CSM mass and the shape of the profile converges at $t>2t_\mathrm{dyn}$, unless $t_\mathrm{dura}\sim t_\mathrm{dyn}$, which justifies the stopping the calculation at $t=2t_\mathrm{dyn}$ in our work. On the other hand, $M_\mathrm{CSM}$ at $t=10t_\mathrm{dyn}$ is slightly larger than that at $t=2t_\mathrm{dyn}$ for models with duration comparable to $t_\mathrm{dyn}$. This may be due to the longer time required for the injected energy to (partially) be converted to kinetic energy for the long-duration models.

For the models with significant mass-loss due to eruption, the profile at $t=10t_{\rm dyn}$ relaxes to the double power-law profile in equation (\ref{eq:rho_CSM_analytical}). This justifies our assumption to use equation (\ref{eq:rho_CSM_analytical}) for the CSM profile at $t=t_{\rm inj}\gg 10t_{\rm dyn}$ when calculating the light curves of the interacting SN.

\bibliography{CSM}{}

\begin{thebibliography}{}
\expandafter\ifx\csname natexlab\endcsname\relax\def\natexlab#1{#1}\fi
\providecommand{\url}[1]{\href{#1}{#1}}
\providecommand{\dodoi}[1]{doi:~\href{http://doi.org/#1}{\nolinkurl{#1}}}
\providecommand{\doeprint}[1]{\href{http://ascl.net/#1}{\nolinkurl{http://ascl.net/#1}}}
\providecommand{\doarXiv}[1]{\href{https://arxiv.org/abs/#1}{\nolinkurl{https://arxiv.org/abs/#1}}}

\bibitem[{{Aamer} {et~al.}(2023){Aamer}, {Nicholl}, {Jerkstrand}, {Gomez},
  {Oates}, {Smartt}, {Srivastav}, {Leloudas}, {Anderson}, {Berger}, {de Boer},
  {Chambers}, {Chen}, {Galbany}, {Gao}, {Gompertz},
  {Gonz{\'a}lez-Ba{\~n}uelos}, {Gromadzki}, {Guti{\'e}rrez}, {Inserra}, {Lowe},
  {Magnier}, {Mazzali}, {Moore}, {M{\"u}ller-Bravo}, {Pursiainen}, {Rest},
  {Schulze}, {Smith}, {Terwel}, {Wainscoat}, \& {Young}}]{Aamer_et_al_2023}
{Aamer}, A., {Nicholl}, M., {Jerkstrand}, A., {et~al.} 2023, arXiv e-prints,
  arXiv:2307.02487, \dodoi{10.48550/arXiv.2307.02487}

\bibitem[{{Arnett}(1982)}]{Arnett82}
{Arnett}, W.~D. 1982, \apj, 253, 785, \dodoi{10.1086/159681}

\bibitem[{{Beasor} {et~al.}(2020){Beasor}, {Davies}, {Smith}, {van Loon},
  {Gehrz}, \& {Figer}}]{Beasor20}
{Beasor}, E.~R., {Davies}, B., {Smith}, N., {et~al.} 2020, \mnras, 492, 5994,
  \dodoi{10.1093/mnras/staa255}

\bibitem[{{Bruch} {et~al.}(2021){Bruch}, {Gal-Yam}, {Schulze}, {Yaron}, {Yang},
  {Soumagnac}, {Rigault}, {Strotjohann}, {Ofek}, {Sollerman}, {Masci},
  {Barbarino}, {Ho}, {Fremling}, {Perley}, {Nordin}, {Cenko}, {Adams},
  {Adreoni}, {Bellm}, {Blagorodnova}, {Bulla}, {Burdge}, {De}, {Dhawan},
  {Drake}, {Duev}, {Dugas}, {Graham}, {Graham}, {Irani}, {Jencson},
  {Karamehmetoglu}, {Kasliwal}, {Kim}, {Kulkarni}, {Kupfer}, {Liang},
  {Mahabal}, {Miller}, {Prince}, {Riddle}, {Sharma}, {Smith}, {Taddia},
  {Taggart}, {Walters}, \& {Yan}}]{Bruch_2021}
{Bruch}, R.~J., {Gal-Yam}, A., {Schulze}, S., {et~al.} 2021, \apj, 912, 46,
  \dodoi{10.3847/1538-4357/abef05}

\bibitem[{{Chevalier}(2012)}]{Chevalier_2012}
{Chevalier}, R.~A. 2012, \apjl, 752, L2, \dodoi{10.1088/2041-8205/752/1/L2}

\bibitem[{{Chevalier} \& {Irwin}(2011)}]{Chevalier_Irwin_2011}
{Chevalier}, R.~A., \& {Irwin}, C.~M. 2011, \apjl, 729, L6,
  \dodoi{10.1088/2041-8205/729/1/L6}

\bibitem[{{Chevalier} \& {Irwin}(2012)}]{Chevalier12}
---. 2012, \apjl, 747, L17, \dodoi{10.1088/2041-8205/747/1/L17}

\bibitem[{{Clocchiatti} \& {Wheeler}(1997)}]{Clocchiatti_Wheeler_1997}
{Clocchiatti}, A., \& {Wheeler}, J.~C. 1997, \apj, 491, 375,
  \dodoi{10.1086/304961}

\bibitem[{{Cold} \& {Hjorth}(2023)}]{Cold23}
{Cold}, C., \& {Hjorth}, J. 2023, \aap, 670, A48,
  \dodoi{10.1051/0004-6361/202244867}

\bibitem[{{Corsi} {et~al.}(2014){Corsi}, {Ofek}, {Gal-Yam}, {Frail},
  {Kulkarni}, {Fox}, {Kasliwal}, {Sullivan}, {Horesh}, {Carpenter}, {Maguire},
  {Arcavi}, {Cenko}, {Cao}, {Mooley}, {Pan}, {Sesar}, {Sternberg}, {Xu},
  {Bersier}, {James}, {Bloom}, \& {Nugent}}]{Corsi14}
{Corsi}, A., {Ofek}, E.~O., {Gal-Yam}, A., {et~al.} 2014, \apj, 782, 42,
  \dodoi{10.1088/0004-637X/782/1/42}

\bibitem[{{Crowther}(2007)}]{Crowther_2007}
{Crowther}, P.~A. 2007, \araa, 45, 177,
  \dodoi{10.1146/annurev.astro.45.051806.110615}

\bibitem[{{Das} {et~al.}(2023){Das}, {Kasliwal}, {Sollerman}, {Fremling},
  {Irani}, {Leung}, {Yang}, {Wu}, {Fuller}, {Anand}, {Andreoni}, {Barbarino},
  {Brink}, {De}, {Dugas}, {Groom}, {Helou}, {Hinds}, {Ho}, {Karambelkar},
  {Kulkarni}, {Perley}, {Purdum}, {Regnault}, {Schulze}, {Sharma}, {Sit},
  {Srinivasaragavan}, {Stein}, {Taggart}, {Tartaglia}, {Tzanidakis}, {Wold},
  {Yan}, {Yao}, \& {Zolkower}}]{Das23}
{Das}, K.~K., {Kasliwal}, M.~M., {Sollerman}, J., {et~al.} 2023, arXiv
  e-prints, arXiv:2306.04698, \dodoi{10.48550/arXiv.2306.04698}

\bibitem[{{de Jager} {et~al.}(1988){de Jager}, {Nieuwenhuijzen}, \& {van der
  Hucht}}]{DeJager88}
{de Jager}, C., {Nieuwenhuijzen}, H., \& {van der Hucht}, K.~A. 1988, \aaps,
  72, 259

\bibitem[{{Dessart} {et~al.}(2022){Dessart}, {Hillier}, \&
  {Kuncarayakti}}]{Dessart_et_al_2022}
{Dessart}, L., {Hillier}, D.~J., \& {Kuncarayakti}, H. 2022, \aap, 658, A130,
  \dodoi{10.1051/0004-6361/202142436}

\bibitem[{{Dong} {et~al.}(2023){Dong}, {Sand}, {Valenti}, {Bostroem},
  {Andrews}, {Hosseinzadeh}, {Hoang}, {Janzen}, {Jencson}, {Lundquist}, {Meza
  Retamal}, {Pearson}, {Shrestha}, {Haislip}, {Kouprianov}, \&
  {Reichart}}]{Dong23}
{Dong}, Y., {Sand}, D.~J., {Valenti}, S., {et~al.} 2023, arXiv e-prints,
  arXiv:2307.02539, \dodoi{10.48550/arXiv.2307.02539}

\bibitem[{{Drout} {et~al.}(2011){Drout}, {Soderberg}, {Gal-Yam}, {Cenko},
  {Fox}, {Leonard}, {Sand}, {Moon}, {Arcavi}, \& {Green}}]{Drout11}
{Drout}, M.~R., {Soderberg}, A.~M., {Gal-Yam}, A., {et~al.} 2011, \apj, 741,
  97, \dodoi{10.1088/0004-637X/741/2/97}

\bibitem[{{Dwek}(1985)}]{Dwek1985}
{Dwek}, E. 1985, \apj, 297, 719, \dodoi{10.1086/163571}

\bibitem[{{Eldridge} {et~al.}(2008){Eldridge}, {Izzard}, \&
  {Tout}}]{Eldridge08}
{Eldridge}, J.~J., {Izzard}, R.~G., \& {Tout}, C.~A. 2008, \mnras, 384, 1109,
  \dodoi{10.1111/j.1365-2966.2007.12738.x}

\bibitem[{{Filippenko}(1997)}]{Filippenko_et_al_1997}
{Filippenko}, A.~V. 1997, \araa, 35, 309,
  \dodoi{10.1146/annurev.astro.35.1.309}

\bibitem[{{F{\"o}rster} {et~al.}(2018){F{\"o}rster}, {Moriya}, {Maureira},
  {Anderson}, {Blinnikov}, {Bufano}, {Cabrera-Vives}, {Clocchiatti}, {de
  Jaeger}, {Est{\'e}vez}, {Galbany}, {Gonz{\'a}lez-Gait{\'a}n}, {Gr{\"a}fener},
  {Hamuy}, {Hsiao}, {Huentelemu}, {Huijse}, {Kuncarayakti}, {Mart{\'\i}nez},
  {Medina}, {Olivares E.}, {Pignata}, {Razza}, {Reyes}, {San Mart{\'\i}n},
  {Smith}, {Vera}, {Vivas}, {de Ugarte Postigo}, {Yoon}, {Ashall}, {Fraser},
  {Gal-Yam}, {Kankare}, {Le Guillou}, {Mazzali}, {Walton}, \&
  {Young}}]{Foerster_et_al_2018}
{F{\"o}rster}, F., {Moriya}, T.~J., {Maureira}, J.~C., {et~al.} 2018, Nature
  Astronomy, 2, 808, \dodoi{10.1038/s41550-018-0563-4}

\bibitem[{{Fryer} \& {Woosley}(1998)}]{Fryer98}
{Fryer}, C.~L., \& {Woosley}, S.~E. 1998, \apjl, 502, L9,
  \dodoi{10.1086/311493}

\bibitem[{{Fuller}(2017)}]{Fuller_2017}
{Fuller}, J. 2017, \mnras, 470, 1642, \dodoi{10.1093/mnras/stx1314}

\bibitem[{{Fuller} \& {Ro}(2018)}]{Fuller_Ro_2018}
{Fuller}, J., \& {Ro}, S. 2018, \mnras, 476, 1853, \dodoi{10.1093/mnras/sty369}

\bibitem[{{Gal-Yam} {et~al.}(2022){Gal-Yam}, {Bruch}, {Schulze}, {Yang},
  {Perley}, {Irani}, {Sollerman}, {Kool}, {Soumagnac}, {Yaron}, {Strotjohann},
  {Zimmerman}, {Barbarino}, {Kulkarni}, {Kasliwal}, {De}, {Yao}, {Fremling},
  {Yan}, {Ofek}, {Fransson}, {Filippenko}, {Zheng}, {Brink}, {Copperwheat},
  {Foley}, {Brown}, {Siebert}, {Leloudas}, {Cabrera-Lavers}, {Garcia-Alvarez},
  {Marante-Barreto}, {Frederick}, {Hung}, {Wheeler}, {Vink{\'o}}, {Thomas},
  {Graham}, {Duev}, {Drake}, {Dekany}, {Bellm}, {Rusholme}, {Shupe},
  {Andreoni}, {Sharma}, {Riddle}, {van Roestel}, \&
  {Knezevic}}]{Gal-Yam_et_al_2022}
{Gal-Yam}, A., {Bruch}, R., {Schulze}, S., {et~al.} 2022, \nat, 601, 201,
  \dodoi{10.1038/s41586-021-04155-1}

\bibitem[{{Gan} {et~al.}(2021){Gan}, {Wang}, \& {Liang}}]{Gan_et_al_2021}
{Gan}, W.-P., {Wang}, S.-Q., \& {Liang}, E.-W. 2021, \apj, 914, 125,
  \dodoi{10.3847/1538-4357/abfbdf}

\bibitem[{{Graur} {et~al.}(2017){Graur}, {Bianco}, {Modjaz}, {Shivvers},
  {Filippenko}, {Li}, \& {Smith}}]{Graur17}
{Graur}, O., {Bianco}, F.~B., {Modjaz}, M., {et~al.} 2017, \apj, 837, 121,
  \dodoi{10.3847/1538-4357/aa5eb7}

\bibitem[{{Hachinger} {et~al.}(2012){Hachinger}, {Mazzali}, {Taubenberger},
  {Hillebrandt}, {Nomoto}, \& {Sauer}}]{Hachinger12}
{Hachinger}, S., {Mazzali}, P.~A., {Taubenberger}, S., {et~al.} 2012, \mnras,
  422, 70, \dodoi{10.1111/j.1365-2966.2012.20464.x}

\bibitem[{Harris {et~al.}(2020)Harris, Millman, van~der Walt, Gommers,
  Virtanen, Cournapeau, Wieser, Taylor, Berg, Smith, Kern, Picus, Hoyer, van
  Kerkwijk, Brett, Haldane, del R{\'{i}}o, Wiebe, Peterson,
  G{\'{e}}rard-Marchant, Sheppard, Reddy, Weckesser, Abbasi, Gohlke, \&
  Oliphant}]{harris2020array}
Harris, C.~R., Millman, K.~J., van~der Walt, S.~J., {et~al.} 2020, Nature, 585,
  357, \dodoi{10.1038/s41586-020-2649-2}

\bibitem[{{Haynie} \& {Piro}(2023)}]{Haynie23}
{Haynie}, A., \& {Piro}, A.~L. 2023, arXiv e-prints, arXiv:2305.12005,
  \dodoi{10.48550/arXiv.2305.12005}

\bibitem[{{Heger} {et~al.}(2003){Heger}, {Fryer}, {Woosley}, {Langer}, \&
  {Hartmann}}]{Heger_et_al_2003}
{Heger}, A., {Fryer}, C.~L., {Woosley}, S.~E., {Langer}, N., \& {Hartmann},
  D.~H. 2003, \apj, 591, 288, \dodoi{10.1086/375341}

\bibitem[{{Heger} \& {Woosley}(2002)}]{Heger02}
{Heger}, A., \& {Woosley}, S.~E. 2002, \apj, 567, 532, \dodoi{10.1086/338487}

\bibitem[{{Hiramatsu} {et~al.}(2023{\natexlab{a}}){Hiramatsu}, {Matsumoto},
  {Berger}, {Ransome}, {Villar}, {Gomez}, {Cendes}, {De}, {Farah}, {Howell},
  {McCully}, {Newsome}, {Padilla Gonzalez}, {Pellegrino}, {Suzuki}, \&
  {Terreran}}]{Hiramatsu23_2021qqp}
{Hiramatsu}, D., {Matsumoto}, T., {Berger}, E., {et~al.} 2023{\natexlab{a}},
  arXiv e-prints, arXiv:2305.11168, \dodoi{10.48550/arXiv.2305.11168}

\bibitem[{{Hiramatsu} {et~al.}(2023{\natexlab{b}}){Hiramatsu}, {Tsuna},
  {Berger}, {Itagaki}, {Goldberg}, {Gomez}, {De}, {Hosseinzadeh}, {Bostroem},
  {Brown}, {Arcavi}, {Bieryla}, {Blanchard}, {Esquerdo}, {Farah}, {Howell},
  {Matsumoto}, {McCully}, {Newsome}, {Padilla Gonzalez}, {Pellegrino}, {Rhee},
  {Terreran}, {Vink{\'o}}, \& {Wheeler}}]{Hiramatsu23}
{Hiramatsu}, D., {Tsuna}, D., {Berger}, E., {et~al.} 2023{\natexlab{b}}, arXiv
  e-prints, arXiv:2307.03165, \dodoi{10.48550/arXiv.2307.03165}

\bibitem[{{Ho} {et~al.}(2019){Ho}, {Goldstein}, {Schulze}, {Khatami}, {Perley},
  {Ergon}, {Gal-Yam}, {Corsi}, {Andreoni}, {Barbarino}, {Bellm},
  {Blagorodnova}, {Bright}, {Burns}, {Cenko}, {Cunningham}, {De}, {Dekany},
  {Dugas}, {Fender}, {Fransson}, {Fremling}, {Goldstein}, {Graham}, {Hale},
  {Horesh}, {Hung}, {Kasliwal}, {Kuin}, {Kulkarni}, {Kupfer}, {Lunnan},
  {Masci}, {Ngeow}, {Nugent}, {Ofek}, {Patterson}, {Petitpas}, {Rusholme},
  {Sai}, {Sfaradi}, {Shupe}, {Sollerman}, {Soumagnac}, {Tachibana}, {Taddia},
  {Walters}, {Wang}, {Yao}, \& {Zhang}}]{Ho19}
{Ho}, A. Y.~Q., {Goldstein}, D.~A., {Schulze}, S., {et~al.} 2019, \apj, 887,
  169, \dodoi{10.3847/1538-4357/ab55ec}

\bibitem[{{Ho} {et~al.}(2023){Ho}, {Perley}, {Gal-Yam}, {Lunnan}, {Sollerman},
  {Schulze}, {Das}, {Dobie}, {Yao}, {Fremling}, {Adams}, {Anand}, {Andreoni},
  {Bellm}, {Bruch}, {Burdge}, {Castro-Tirado}, {Dahiwale}, {De}, {Dekany},
  {Drake}, {Duev}, {Graham}, {Helou}, {Kaplan}, {Karambelkar}, {Kasliwal},
  {Kool}, {Kulkarni}, {Mahabal}, {Medford}, {Miller}, {Nordin}, {Ofek},
  {Petitpas}, {Riddle}, {Sharma}, {Smith}, {Stewart}, {Taggart}, {Tartaglia},
  {Tzanidakis}, \& {Winters}}]{Ho23}
{Ho}, A. Y.~Q., {Perley}, D.~A., {Gal-Yam}, A., {et~al.} 2023, \apj, 949, 120,
  \dodoi{10.3847/1538-4357/acc533}

\bibitem[{{Hosseinzadeh} {et~al.}(2019){Hosseinzadeh}, {McCully}, {Zabludoff},
  {Arcavi}, {French}, {Howell}, {Berger}, \&
  {Hiramatsu}}]{Hosseinzadeh_et_al_2019}
{Hosseinzadeh}, G., {McCully}, C., {Zabludoff}, A.~I., {et~al.} 2019, \apjl,
  871, L9, \dodoi{10.3847/2041-8213/aafc61}

\bibitem[{{Hosseinzadeh} {et~al.}(2017){Hosseinzadeh}, {Arcavi}, {Valenti},
  {McCully}, {Howell}, {Johansson}, {Sollerman}, {Pastorello}, {Benetti},
  {Cao}, {Cenko}, {Clubb}, {Corsi}, {Duggan}, {Elias-Rosa}, {Filippenko},
  {Fox}, {Fremling}, {Horesh}, {Karamehmetoglu}, {Kasliwal}, {Marion}, {Ofek},
  {Sand}, {Taddia}, {Zheng}, {Fraser}, {Gal-Yam}, {Inserra}, {Laher}, {Masci},
  {Rebbapragada}, {Smartt}, {Smith}, {Sullivan}, {Surace}, \&
  {Wo{\'z}niak}}]{Hosseinzadeh_et_al_2017}
{Hosseinzadeh}, G., {Arcavi}, I., {Valenti}, S., {et~al.} 2017, \apj, 836, 158,
  \dodoi{10.3847/1538-4357/836/2/158}

\bibitem[{Hunter(2007)}]{Hunter:2007}
Hunter, J.~D. 2007, Computing in Science \& Engineering, 9, 90,
  \dodoi{10.1109/MCSE.2007.55}

\bibitem[{{Iglesias} \& {Rogers}(1996)}]{Ross}
{Iglesias}, C.~A., \& {Rogers}, F.~J. 1996, \apj, 464, 943,
  \dodoi{10.1086/177381}

\bibitem[{{Ivezi{\'c}} {et~al.}(2019){Ivezi{\'c}}, {Kahn}, {Tyson}, {Abel},
  {Acosta}, {Allsman}, {Alonso}, {AlSayyad}, {Anderson}, {Andrew}, {Angel},
  {Angeli}, {Ansari}, {Antilogus}, {Araujo}, {Armstrong}, {Arndt}, {Astier},
  {Aubourg}, {Auza}, {Axelrod}, {Bard}, {Barr}, {Barrau}, {Bartlett}, {Bauer},
  {Bauman}, {Baumont}, {Bechtol}, {Bechtol}, {Becker}, {Becla}, {Beldica},
  {Bellavia}, {Bianco}, {Biswas}, {Blanc}, {Blazek}, {Blandford}, {Bloom},
  {Bogart}, {Bond}, {Booth}, {Borgland}, {Borne}, {Bosch}, {Boutigny},
  {Brackett}, {Bradshaw}, {Brandt}, {Brown}, {Bullock}, {Burchat}, {Burke},
  {Cagnoli}, {Calabrese}, {Callahan}, {Callen}, {Carlin}, {Carlson},
  {Chandrasekharan}, {Charles-Emerson}, {Chesley}, {Cheu}, {Chiang}, {Chiang},
  {Chirino}, {Chow}, {Ciardi}, {Claver}, {Cohen-Tanugi}, {Cockrum}, {Coles},
  {Connolly}, {Cook}, {Cooray}, {Covey}, {Cribbs}, {Cui}, {Cutri}, {Daly},
  {Daniel}, {Daruich}, {Daubard}, {Daues}, {Dawson}, {Delgado}, {Dellapenna},
  {de Peyster}, {de Val-Borro}, {Digel}, {Doherty}, {Dubois},
  {Dubois-Felsmann}, {Durech}, {Economou}, {Eifler}, {Eracleous}, {Emmons},
  {Fausti Neto}, {Ferguson}, {Figueroa}, {Fisher-Levine}, {Focke}, {Foss},
  {Frank}, {Freemon}, {Gangler}, {Gawiser}, {Geary}, {Gee}, {Geha}, {Gessner},
  {Gibson}, {Gilmore}, {Glanzman}, {Glick}, {Goldina}, {Goldstein}, {Goodenow},
  {Graham}, {Gressler}, {Gris}, {Guy}, {Guyonnet}, {Haller}, {Harris},
  {Hascall}, {Haupt}, {Hernandez}, {Herrmann}, {Hileman}, {Hoblitt}, {Hodgson},
  {Hogan}, {Howard}, {Huang}, {Huffer}, {Ingraham}, {Innes}, {Jacoby}, {Jain},
  {Jammes}, {Jee}, {Jenness}, {Jernigan}, {Jevremovi{\'c}}, {Johns}, {Johnson},
  {Johnson}, {Jones}, {Juramy-Gilles}, {Juri{\'c}}, {Kalirai}, {Kallivayalil},
  {Kalmbach}, {Kantor}, {Karst}, {Kasliwal}, {Kelly}, {Kessler}, {Kinnison},
  {Kirkby}, {Knox}, {Kotov}, {Krabbendam}, {Krughoff}, {Kub{\'a}nek},
  {Kuczewski}, {Kulkarni}, {Ku}, {Kurita}, {Lage}, {Lambert}, {Lange},
  {Langton}, {Le Guillou}, {Levine}, {Liang}, {Lim}, {Lintott}, {Long},
  {Lopez}, {Lotz}, {Lupton}, {Lust}, {MacArthur}, {Mahabal}, {Mandelbaum},
  {Markiewicz}, {Marsh}, {Marshall}, {Marshall}, {May}, {McKercher}, {McQueen},
  {Meyers}, {Migliore}, {Miller}, {Mills}, {Miraval}, {Moeyens}, {Moolekamp},
  {Monet}, {Moniez}, {Monkewitz}, {Montgomery}, {Morrison}, {Mueller},
  {Muller}, {Mu{\~n}oz Arancibia}, {Neill}, {Newbry}, {Nief}, {Nomerotski},
  {Nordby}, {O'Connor}, {Oliver}, {Olivier}, {Olsen}, {O'Mullane}, {Ortiz},
  {Osier}, {Owen}, {Pain}, {Palecek}, {Parejko}, {Parsons}, {Pease},
  {Peterson}, {Peterson}, {Petravick}, {Libby Petrick}, {Petry},
  {Pierfederici}, {Pietrowicz}, {Pike}, {Pinto}, {Plante}, {Plate}, {Plutchak},
  {Price}, {Prouza}, {Radeka}, {Rajagopal}, {Rasmussen}, {Regnault}, {Reil},
  {Reiss}, {Reuter}, {Ridgway}, {Riot}, {Ritz}, {Robinson}, {Roby}, {Roodman},
  {Rosing}, {Roucelle}, {Rumore}, {Russo}, {Saha}, {Sassolas}, {Schalk},
  {Schellart}, {Schindler}, {Schmidt}, {Schneider}, {Schneider}, {Schoening},
  {Schumacher}, {Schwamb}, {Sebag}, {Selvy}, {Sembroski}, {Seppala}, {Serio},
  {Serrano}, {Shaw}, {Shipsey}, {Sick}, {Silvestri}, {Slater}, {Smith},
  {Smith}, {Sobhani}, {Soldahl}, {Storrie-Lombardi}, {Stover}, {Strauss},
  {Street}, {Stubbs}, {Sullivan}, {Sweeney}, {Swinbank}, {Szalay}, {Takacs},
  {Tether}, {Thaler}, {Thayer}, {Thomas}, {Thornton}, {Thukral}, {Tice},
  {Trilling}, {Turri}, {Van Berg}, {Vanden Berk}, {Vetter}, {Virieux},
  {Vucina}, {Wahl}, {Walkowicz}, {Walsh}, {Walter}, {Wang}, {Wang}, {Warner},
  {Wiecha}, {Willman}, {Winters}, {Wittman}, {Wolff}, {Wood-Vasey}, {Wu},
  {Xin}, {Yoachim}, \& {Zhan}}]{LSST_2019}
{Ivezi{\'c}}, {\v{Z}}., {Kahn}, S.~M., {Tyson}, J.~A., {et~al.} 2019, \apj,
  873, 111, \dodoi{10.3847/1538-4357/ab042c}

\bibitem[{{Jacobson-Gal{\'a}n} {et~al.}(2022){Jacobson-Gal{\'a}n}, {Dessart},
  {Jones}, {Margutti}, {Coppejans}, {Dimitriadis}, {Foley}, {Kilpatrick},
  {Matthews}, {Rest}, {Terreran}, {Aleo}, {Auchettl}, {Blanchard}, {Coulter},
  {Davis}, {de Boer}, {DeMarchi}, {Drout}, {Earl}, {Gagliano}, {Gall},
  {Hjorth}, {Huber}, {Ibik}, {Milisavljevic}, {Pan}, {Rest}, {Ridden-Harper},
  {Rojas-Bravo}, {Siebert}, {Smith}, {Taggart}, {Tinyanont}, {Wang}, \&
  {Zenati}}]{Jacobson_et_al_2022}
{Jacobson-Gal{\'a}n}, W.~V., {Dessart}, L., {Jones}, D.~O., {et~al.} 2022,
  \apj, 924, 15, \dodoi{10.3847/1538-4357/ac3f3a}

\bibitem[{{Jermyn} {et~al.}(2023){Jermyn}, {Bauer}, {Schwab}, {Farmer}, {Ball},
  {Bellinger}, {Dotter}, {Joyce}, {Marchant}, {Mombarg}, {Wolf}, {Sunny Wong},
  {Cinquegrana}, {Farrell}, {Smolec}, {Thoul}, {Cantiello}, {Herwig}, {Toloza},
  {Bildsten}, {Townsend}, \& {Timmes}}]{Jermyn23}
{Jermyn}, A.~S., {Bauer}, E.~B., {Schwab}, J., {et~al.} 2023, \apjs, 265, 15,
  \dodoi{10.3847/1538-4365/acae8d}

\bibitem[{{Khazov} {et~al.}(2016){Khazov}, {Yaron}, {Gal-Yam}, {Manulis},
  {Rubin}, {Kulkarni}, {Arcavi}, {Kasliwal}, {Ofek}, {Cao}, {Perley},
  {Sollerman}, {Horesh}, {Sullivan}, {Filippenko}, {Nugent}, {Howell}, {Cenko},
  {Silverman}, {Ebeling}, {Taddia}, {Johansson}, {Laher}, {Surace},
  {Rebbapragada}, {Wozniak}, \& {Matheson}}]{Khazov_2016}
{Khazov}, D., {Yaron}, O., {Gal-Yam}, A., {et~al.} 2016, \apj, 818, 3,
  \dodoi{10.3847/0004-637X/818/1/3}

\bibitem[{{Kirchschlager} {et~al.}(2019){Kirchschlager}, {Schmidt}, {Barlow},
  {Fogerty}, {Bevan}, \& {Priestley}}]{Kirchschlager19}
{Kirchschlager}, F., {Schmidt}, F.~D., {Barlow}, M.~J., {et~al.} 2019, \mnras,
  489, 4465, \dodoi{10.1093/mnras/stz2399}

\bibitem[{{Ko} {et~al.}(2022){Ko}, {Tsuna}, {Takei}, \&
  {Shigeyama}}]{Ko_et_al_2022}
{Ko}, T., {Tsuna}, D., {Takei}, Y., \& {Shigeyama}, T. 2022, \apj, 930, 168,
  \dodoi{10.3847/1538-4357/ac67e1}

\bibitem[{{Kuncarayakti} {et~al.}(2022){Kuncarayakti}, {Maeda}, {Dessart},
  {Nagao}, {Fulton}, {Guti{\'e}rrez}, {Huber}, {Young}, {Kotak}, {Mattila},
  {Anderson}, {Ferrari}, {Folatelli}, {Gao}, {Magnier}, {Smith}, \&
  {Srivastav}}]{Kuncarayakti22}
{Kuncarayakti}, H., {Maeda}, K., {Dessart}, L., {et~al.} 2022, \apjl, 941, L32,
  \dodoi{10.3847/2041-8213/aca672}

\bibitem[{{Kuncarayakti} {et~al.}(2023){Kuncarayakti}, {Sollerman}, {Izzo},
  {Maeda}, {Yang}, {Schulze}, {Angus}, {Aubert}, {Auchettl}, {Della Valle},
  {Dessart}, {Hinds}, {Kankare}, {Kawabata}, {Lundqvist}, {Nakaoka}, {Perley},
  {Raimundo}, {Strotjohann}, {Taguchi}, {Cai}, {Charalampopoulos}, {Fang},
  {Fraser}, {Gutierrez}, {Imazawa}, {Kangas}, {Kawabata}, {Kotak}, {Kravtsov},
  {Matilainen}, {Mattila}, {Moran}, {Murata}, {Salmaso}, {Anderson}, {Ashall},
  {Bellm}, {Benetti}, {Chambers}, {Chen}, {Coughlin}, {De Colle}, {Fremling},
  {Galbany}, {Gal-Yam}, {Gromadzki}, {Groom}, {Hajela}, {Inserra}, {Kasliwal},
  {Mahabal}, {Martin-Carrillo}, {Moore}, {Muller-Bravo}, {Nicholl}, {Ragosta},
  {Riddle}, {Sharma}, {Srivastav}, {Stritzinger}, {Wold}, \&
  {Young}}]{Kuncarayakti_2023}
{Kuncarayakti}, H., {Sollerman}, J., {Izzo}, L., {et~al.} 2023, arXiv e-prints,
  arXiv:2303.16925, \dodoi{10.48550/arXiv.2303.16925}

\bibitem[{{Kuriyama} \& {Shigeyama}(2020)}]{Kuriyama20a}
{Kuriyama}, N., \& {Shigeyama}, T. 2020, \aap, 635, A127,
  \dodoi{10.1051/0004-6361/201937226}

\bibitem[{{Kuriyama} \& {Shigeyama}(2021)}]{Kuriyama21}
---. 2021, \aap, 646, A118, \dodoi{10.1051/0004-6361/202038637}

\bibitem[{{Leung} {et~al.}(2019){Leung}, {Nomoto}, \& {Blinnikov}}]{Leung19}
{Leung}, S.-C., {Nomoto}, K., \& {Blinnikov}, S. 2019, \apj, 887, 72,
  \dodoi{10.3847/1538-4357/ab4fe5}

\bibitem[{{Leung} {et~al.}(2021){Leung}, {Wu}, \& {Fuller}}]{Leung_et_al_2021}
{Leung}, S.-C., {Wu}, S., \& {Fuller}, J. 2021, \apj, 923, 41,
  \dodoi{10.3847/1538-4357/ac2c63}

\bibitem[{{Lyman} {et~al.}(2016){Lyman}, {Bersier}, {James}, {Mazzali},
  {Eldridge}, {Fraser}, \& {Pian}}]{Lyman_et_al_2016}
{Lyman}, J.~D., {Bersier}, D., {James}, P.~A., {et~al.} 2016, \mnras, 457, 328,
  \dodoi{10.1093/mnras/stv2983}

\bibitem[{{Maeda} \& {Moriya}(2022)}]{Maeda_Moriya_2022}
{Maeda}, K., \& {Moriya}, T.~J. 2022, \apj, 927, 25,
  \dodoi{10.3847/1538-4357/ac4672}

\bibitem[{{Magee} {et~al.}(1995){Magee}, {Abdallah}, {Clark}, {Cohen},
  {Collins}, {Csanak}, {Fontes}, {Gauger}, {Keady}, {Kilcrease}, \&
  {Merts}}]{1995ASPC...78...51M}
{Magee}, N.~H., {Abdallah}, J., J., {Clark}, R.~E.~H., {et~al.} 1995, in
  Astronomical Society of the Pacific Conference Series, Vol.~78, Astrophysical
  Applications of Powerful New Databases, ed. S.~J. {Adelman} \& W.~L. {Wiese},
  51

\bibitem[{{Margutti} {et~al.}(2014){Margutti}, {Milisavljevic}, {Soderberg},
  {Chornock}, {Zauderer}, {Murase}, {Guidorzi}, {Sanders}, {Kuin}, {Fransson},
  {Levesque}, {Chandra}, {Berger}, {Bianco}, {Brown}, {Challis},
  {Chatzopoulos}, {Cheung}, {Choi}, {Chomiuk}, {Chugai}, {Contreras}, {Drout},
  {Fesen}, {Foley}, {Fong}, {Friedman}, {Gall}, {Gehrels}, {Hjorth}, {Hsiao},
  {Kirshner}, {Im}, {Leloudas}, {Lunnan}, {Marion}, {Martin}, {Morrell},
  {Neugent}, {Omodei}, {Phillips}, {Rest}, {Silverman}, {Strader},
  {Stritzinger}, {Szalai}, {Utterback}, {Vinko}, {Wheeler}, {Arnett},
  {Campana}, {Chevalier}, {Ginsburg}, {Kamble}, {Roming}, {Pritchard}, \&
  {Stringfellow}}]{Margutti_et_al_2014}
{Margutti}, R., {Milisavljevic}, D., {Soderberg}, A.~M., {et~al.} 2014, \apj,
  780, 21, \dodoi{10.1088/0004-637X/780/1/21}

\bibitem[{{Marigo} \& {Aringer}(2009)}]{Marigo_Aringer_2009}
{Marigo}, P., \& {Aringer}, B. 2009, \aap, 508, 1539,
  \dodoi{10.1051/0004-6361/200912598}

\bibitem[{{Marigo} {et~al.}(2022){Marigo}, {Aringer}, {Girardi}, \&
  {Bressan}}]{Marigo_et_al_2022}
{Marigo}, P., {Aringer}, B., {Girardi}, L., \& {Bressan}, A. 2022, \apj, 940,
  129, \dodoi{10.3847/1538-4357/ac9b40}

\bibitem[{{Matheson} {et~al.}(2000){Matheson}, {Filippenko}, {Chornock},
  {Leonard}, \& {Li}}]{Matheson_et_al_2000}
{Matheson}, T., {Filippenko}, A.~V., {Chornock}, R., {Leonard}, D.~C., \& {Li},
  W. 2000, \aj, 119, 2303, \dodoi{10.1086/301352}

\bibitem[{{Mattila} {et~al.}(2008){Mattila}, {Meikle}, {Lundqvist},
  {Pastorello}, {Kotak}, {Eldridge}, {Smartt}, {Adamson}, {Gerardy}, {Rizzi},
  {Stephens}, \& {van Dyk}}]{Mattila_et_al_2008}
{Mattila}, S., {Meikle}, W.~P.~S., {Lundqvist}, P., {et~al.} 2008, \mnras, 389,
  141, \dodoi{10.1111/j.1365-2966.2008.13516.x}

\bibitem[{{Matzner} \& {McKee}(1999)}]{Matzner99}
{Matzner}, C.~D., \& {McKee}, C.~F. 1999, \apj, 510, 379,
  \dodoi{10.1086/306571}

\bibitem[{{Mauron} \& {Josselin}(2011)}]{Mauron11}
{Mauron}, N., \& {Josselin}, E. 2011, \aap, 526, A156,
  \dodoi{10.1051/0004-6361/201013993}

\bibitem[{{Metzger}(2022)}]{Metzger_2022}
{Metzger}, B.~D. 2022, \apj, 932, 84, \dodoi{10.3847/1538-4357/ac6d59}

\bibitem[{{Moriya} {et~al.}(2011){Moriya}, {Tominaga}, {Blinnikov}, {Baklanov},
  \& {Sorokina}}]{Moriya_et_al_2011}
{Moriya}, T., {Tominaga}, N., {Blinnikov}, S.~I., {Baklanov}, P.~V., \&
  {Sorokina}, E.~I. 2011, \mnras, 415, 199,
  \dodoi{10.1111/j.1365-2966.2011.18689.x}

\bibitem[{{Moriya} \& {Maeda}(2016)}]{Moriya_Maeda_2016}
{Moriya}, T.~J., \& {Maeda}, K. 2016, \apj, 824, 100,
  \dodoi{10.3847/0004-637X/824/2/100}

\bibitem[{{Morozova} {et~al.}(2018){Morozova}, {Piro}, \&
  {Valenti}}]{Morozova18}
{Morozova}, V., {Piro}, A.~L., \& {Valenti}, S. 2018, \apj, 858, 15,
  \dodoi{10.3847/1538-4357/aab9a6}

\bibitem[{{Nakano} {et~al.}(2006){Nakano}, {Itagaki}, {Puckett}, \&
  {Gorelli}}]{Nakano_et_al_2006}
{Nakano}, S., {Itagaki}, K., {Puckett}, T., \& {Gorelli}, R. 2006, Central
  Bureau Electronic Telegrams, 666, 1

\bibitem[{{Nakar}(2015)}]{Nakar15}
{Nakar}, E. 2015, \apj, 807, 172, \dodoi{10.1088/0004-637X/807/2/172}

\bibitem[{{Nakar} \& {Piro}(2014)}]{Nakar_Piro_14}
{Nakar}, E., \& {Piro}, A.~L. 2014, \apj, 788, 193,
  \dodoi{10.1088/0004-637X/788/2/193}

\bibitem[{{Nakar} \& {Sari}(2012)}]{Nakar_Sari_2012}
{Nakar}, E., \& {Sari}, R. 2012, \apj, 747, 88,
  \dodoi{10.1088/0004-637X/747/2/88}

\bibitem[{{Neustadt} {et~al.}(2023){Neustadt}, {Kochanek}, \& {Rizzo
  Smith}}]{Neustadt23}
{Neustadt}, J.~M.~M., {Kochanek}, C.~S., \& {Rizzo Smith}, M. 2023, arXiv
  e-prints, arXiv:2306.06162, \dodoi{10.48550/arXiv.2306.06162}

\bibitem[{{Nicholl} {et~al.}(2015){Nicholl}, {Smartt}, {Jerkstrand}, {Sim},
  {Inserra}, {Anderson}, {Baltay}, {Benetti}, {Chambers}, {Chen}, {Elias-Rosa},
  {Feindt}, {Flewelling}, {Fraser}, {Gal-Yam}, {Galbany}, {Huber}, {Kangas},
  {Kankare}, {Kotak}, {Kr{\"u}hler}, {Maguire}, {McKinnon}, {Rabinowitz},
  {Rostami}, {Schulze}, {Smith}, {Sullivan}, {Tonry}, {Valenti}, \&
  {Young}}]{Nicholl_et_al_2015}
{Nicholl}, M., {Smartt}, S.~J., {Jerkstrand}, A., {et~al.} 2015, \apjl, 807,
  L18, \dodoi{10.1088/2041-8205/807/1/L18}

\bibitem[{{Ofek} {et~al.}(2014){Ofek}, {Sullivan}, {Shaviv}, {Steinbok},
  {Arcavi}, {Gal-Yam}, {Tal}, {Kulkarni}, {Nugent}, {Ben-Ami}, {Kasliwal},
  {Cenko}, {Laher}, {Surace}, {Bloom}, {Filippenko}, {Silverman}, \&
  {Yaron}}]{Ofek14}
{Ofek}, E.~O., {Sullivan}, M., {Shaviv}, N.~J., {et~al.} 2014, \apj, 789, 104,
  \dodoi{10.1088/0004-637X/789/2/104}

\bibitem[{{Omand} {et~al.}(2019){Omand}, {Kashiyama}, \& {Murase}}]{Omand19}
{Omand}, C. M.~B., {Kashiyama}, K., \& {Murase}, K. 2019, \mnras, 484, 5468,
  \dodoi{10.1093/mnras/stz371}

\bibitem[{{Pastorello} {et~al.}(2007){Pastorello}, {Smartt}, {Mattila},
  {Eldridge}, {Young}, {Itagaki}, {Yamaoka}, {Navasardyan}, {Valenti}, {Patat},
  {Agnoletto}, {Augusteijn}, {Benetti}, {Cappellaro}, {Boles}, {Bonnet-Bidaud},
  {Botticella}, {Bufano}, {Cao}, {Deng}, {Dennefeld}, {Elias-Rosa},
  {Harutyunyan}, {Keenan}, {Iijima}, {Lorenzi}, {Mazzali}, {Meng}, {Nakano},
  {Nielsen}, {Smoker}, {Stanishev}, {Turatto}, {Xu}, \&
  {Zampieri}}]{Pastorello_et_al_2007}
{Pastorello}, A., {Smartt}, S.~J., {Mattila}, S., {et~al.} 2007, \nat, 447,
  829, \dodoi{10.1038/nature05825}

\bibitem[{{Pastorello} {et~al.}(2008){Pastorello}, {Mattila}, {Zampieri},
  {Della Valle}, {Smartt}, {Valenti}, {Agnoletto}, {Benetti}, {Benn}, {Branch},
  {Cappellaro}, {Dennefeld}, {Eldridge}, {Gal-Yam}, {Harutyunyan}, {Hunter},
  {Kjeldsen}, {Lipkin}, {Mazzali}, {Milne}, {Navasardyan}, {Ofek}, {Pian},
  {Shemmer}, {Spiro}, {Stathakis}, {Taubenberger}, {Turatto}, \&
  {Yamaoka}}]{Pastorello08}
{Pastorello}, A., {Mattila}, S., {Zampieri}, L., {et~al.} 2008, \mnras, 389,
  113, \dodoi{10.1111/j.1365-2966.2008.13602.x}

\bibitem[{{Pastorello} {et~al.}(2015{\natexlab{a}}){Pastorello}, {Hadjiyska},
  {Rabinowitz}, {Valenti}, {Turatto}, {Fasano}, {Benitez-Herrera}, {Baltay},
  {Benetti}, {Botticella}, {Cappellaro}, {Elias-Rosa}, {Ellman}, {Feindt},
  {Filippenko}, {Fraser}, {Gal-Yam}, {Graham}, {Howell}, {Inserra}, {Kelly},
  {Kotak}, {Kowalski}, {McKinnon}, {Morales-Garoffolo}, {Nugent}, {Smartt},
  {Smith}, {Stritzinger}, {Sullivan}, {Taubenberger}, {Walker}, {Yaron}, \&
  {Young}}]{Pastorello_et_al_2015_ccw_btw}
{Pastorello}, A., {Hadjiyska}, E., {Rabinowitz}, D., {et~al.}
  2015{\natexlab{a}}, \mnras, 449, 1954, \dodoi{10.1093/mnras/stv335}

\bibitem[{{Pastorello} {et~al.}(2015{\natexlab{b}}){Pastorello}, {Prieto},
  {Elias-Rosa}, {Bersier}, {Hosseinzadeh}, {Morales-Garoffolo}, {Noebauer},
  {Taubenberger}, {Tomasella}, {Kochanek}, {Falco}, {Basu}, {Beacom},
  {Benetti}, {Brimacombe}, {Cappellaro}, {Danilet}, {Dong}, {Fernandez},
  {Goss}, {Granata}, {Harutyunyan}, {Holoien}, {Ishida}, {Kiyota}, {Krannich},
  {Nicholls}, {Ochner}, {Pojma{\'n}ski}, {Shappee}, {Simonian}, {Stanek},
  {Starrfield}, {Szczygie{\l}}, {Tartaglia}, {Terreran}, {Thompson}, {Turatto},
  {Wagner}, {Wiethoff}, {Wilber}, \&
  {Wo{\'z}niak}}]{Pastorello_2015_ASASSN_15ed}
{Pastorello}, A., {Prieto}, J.~L., {Elias-Rosa}, N., {et~al.}
  2015{\natexlab{b}}, \mnras, 453, 3649, \dodoi{10.1093/mnras/stv1812}

\bibitem[{{Pastorello} {et~al.}(2015{\natexlab{c}}){Pastorello}, {Wyrzykowski},
  {Valenti}, {Prieto}, {Koz{\l}owski}, {Udalski}, {Elias-Rosa},
  {Morales-Garoffolo}, {Anderson}, {Benetti}, {Bersten}, {Botticella},
  {Cappellaro}, {Fasano}, {Fraser}, {Gal-Yam}, {Gillone}, {Graham}, {Greiner},
  {Hachinger}, {Howell}, {Inserra}, {Parrent}, {Rau}, {Schulze}, {Smartt},
  {Smith}, {Turatto}, {Yaron}, {Young}, {Kubiak}, {Szyma{\'n}ski},
  {Pietrzy{\'n}ski}, {Soszy{\'n}ski}, {Ulaczyk}, {Poleski}, {Pietrukowicz},
  {Skowron}, \& {Mr{\'o}z}}]{Pastorello_et_al_2015_OGLE}
{Pastorello}, A., {Wyrzykowski}, {\L}., {Valenti}, S., {et~al.}
  2015{\natexlab{c}}, \mnras, 449, 1941, \dodoi{10.1093/mnras/stu2621}

\bibitem[{{Pastorello} {et~al.}(2016){Pastorello}, {Wang}, {Ciabattari},
  {Bersier}, {Mazzali}, {Gao}, {Xu}, {Zhang}, {Tokuoka}, {Benetti},
  {Cappellaro}, {Elias-Rosa}, {Harutyunyan}, {Huang}, {Miluzio}, {Mo},
  {Ochner}, {Tartaglia}, {Terreran}, {Tomasella}, \&
  {Turatto}}]{Pastorello_2016_2014av}
{Pastorello}, A., {Wang}, X.~F., {Ciabattari}, F., {et~al.} 2016, \mnras, 456,
  853, \dodoi{10.1093/mnras/stv2634}

\bibitem[{{Paxton} {et~al.}(2011){Paxton}, {Bildsten}, {Dotter}, {Herwig},
  {Lesaffre}, \& {Timmes}}]{Paxton11}
{Paxton}, B., {Bildsten}, L., {Dotter}, A., {et~al.} 2011, \apjs, 192, 3,
  \dodoi{10.1088/0067-0049/192/1/3}

\bibitem[{{Paxton} {et~al.}(2013){Paxton}, {Cantiello}, {Arras}, {Bildsten},
  {Brown}, {Dotter}, {Mankovich}, {Montgomery}, {Stello}, {Timmes}, \&
  {Townsend}}]{Paxton13}
{Paxton}, B., {Cantiello}, M., {Arras}, P., {et~al.} 2013, \apjs, 208, 4,
  \dodoi{10.1088/0067-0049/208/1/4}

\bibitem[{{Paxton} {et~al.}(2015){Paxton}, {Marchant}, {Schwab}, {Bauer},
  {Bildsten}, {Cantiello}, {Dessart}, {Farmer}, {Hu}, {Langer}, {Townsend},
  {Townsley}, \& {Timmes}}]{Paxton15}
{Paxton}, B., {Marchant}, P., {Schwab}, J., {et~al.} 2015, \apjs, 220, 15,
  \dodoi{10.1088/0067-0049/220/1/15}

\bibitem[{{Paxton} {et~al.}(2018){Paxton}, {Schwab}, {Bauer}, {Bildsten},
  {Blinnikov}, {Duffell}, {Farmer}, {Goldberg}, {Marchant}, {Sorokina},
  {Thoul}, {Townsend}, \& {Timmes}}]{Paxton18}
{Paxton}, B., {Schwab}, J., {Bauer}, E.~B., {et~al.} 2018, \apjs, 234, 34,
  \dodoi{10.3847/1538-4365/aaa5a8}

\bibitem[{{Paxton} {et~al.}(2019){Paxton}, {Smolec}, {Schwab}, {Gautschy},
  {Bildsten}, {Cantiello}, {Dotter}, {Farmer}, {Goldberg}, {Jermyn}, {Kanbur},
  {Marchant}, {Thoul}, {Townsend}, {Wolf}, {Zhang}, \& {Timmes}}]{Paxton19}
{Paxton}, B., {Smolec}, R., {Schwab}, J., {et~al.} 2019, \apjs, 243, 10,
  \dodoi{10.3847/1538-4365/ab2241}

\bibitem[{{Pellegrino} {et~al.}(2022){Pellegrino}, {Howell}, {Terreran},
  {Arcavi}, {Bostroem}, {Brown}, {Burke}, {Dong}, {Gilkis}, {Hiramatsu},
  {Hosseinzadeh}, {McCully}, {Modjaz}, {Newsome}, {Gonzalez}, {Pritchard},
  {Sand}, {Valenti}, \& {Williamson}}]{Pellegrino22}
{Pellegrino}, C., {Howell}, D.~A., {Terreran}, G., {et~al.} 2022, \apj, 938,
  73, \dodoi{10.3847/1538-4357/ac8ff6}

\bibitem[{{Perley} {et~al.}(2022){Perley}, {Sollerman}, {Schulze}, {Yao},
  {Fremling}, {Gal-Yam}, {Ho}, {Yang}, {Kool}, {Irani}, {Yan}, {Andreoni},
  {Baade}, {Bellm}, {Brink}, {Chen}, {Cikota}, {Coughlin}, {Dahiwale},
  {Dekany}, {Duev}, {Filippenko}, {Hoeflich}, {Kasliwal}, {Kulkarni}, {Lunnan},
  {Masci}, {Maund}, {Medford}, {Riddle}, {Rosnet}, {Shupe}, {Strotjohann},
  {Tzanidakis}, \& {Zheng}}]{Perley22}
{Perley}, D.~A., {Sollerman}, J., {Schulze}, S., {et~al.} 2022, \apj, 927, 180,
  \dodoi{10.3847/1538-4357/ac478e}

\bibitem[{{Podsiadlowski} {et~al.}(1992){Podsiadlowski}, {Joss}, \&
  {Hsu}}]{Podsiadlowski_et_al_1992}
{Podsiadlowski}, P., {Joss}, P.~C., \& {Hsu}, J.~J.~L. 1992, \apj, 391, 246,
  \dodoi{10.1086/171341}

\bibitem[{{Prentice} {et~al.}(2016){Prentice}, {Mazzali}, {Pian}, {Gal-Yam},
  {Kulkarni}, {Rubin}, {Corsi}, {Fremling}, {Sollerman}, {Yaron}, {Arcavi},
  {Zheng}, {Kasliwal}, {Filippenko}, {Cenko}, {Cao}, \& {Nugent}}]{Prentice16}
{Prentice}, S.~J., {Mazzali}, P.~A., {Pian}, E., {et~al.} 2016, \mnras, 458,
  2973, \dodoi{10.1093/mnras/stw299}

\bibitem[{{Pursiainen} {et~al.}(2023){Pursiainen}, {Leloudas}, {Schulze},
  {Charalampopoulos}, {Angus}, {Anderson}, {Bauer}, {Chen}, {Galbany},
  {Gromadzki}, {Guti{\'e}rrez}, {Inserra}, {M{\"u}ller-Bravo}, {Nicholl},
  {Smartt}, {Tartaglia}, {Wiseman}, \& {Young}}]{Pursiainen_2023}
{Pursiainen}, M., {Leloudas}, G., {Schulze}, S., {et~al.} 2023, arXiv e-prints,
  arXiv:2306.09804, \dodoi{10.48550/arXiv.2306.09804}

\bibitem[{{Quataert} \& {Shiode}(2012)}]{2012MNRAS.423L..92Q}
{Quataert}, E., \& {Shiode}, J. 2012, \mnras, 423, L92,
  \dodoi{10.1111/j.1745-3933.2012.01264.x}

\bibitem[{{Renzo} {et~al.}(2020){Renzo}, {Farmer}, {Justham}, {G{\"o}tberg},
  {de Mink}, {Zapartas}, {Marchant}, \& {Smith}}]{Renzo_et_al_2020}
{Renzo}, M., {Farmer}, R., {Justham}, S., {et~al.} 2020, \aap, 640, A56,
  \dodoi{10.1051/0004-6361/202037710}

\bibitem[{{Sarangi} \& {Cherchneff}(2013)}]{Sarangi_2013}
{Sarangi}, A., \& {Cherchneff}, I. 2013, \apj, 776, 107,
  \dodoi{10.1088/0004-637X/776/2/107}

\bibitem[{{Schlegel}(1990)}]{1990MNRAS.244..269S}
{Schlegel}, E.~M. 1990, \mnras, 244, 269

\bibitem[{{Schneider} {et~al.}(2021){Schneider}, {Podsiadlowski}, \&
  {M{\"u}ller}}]{Schneider21}
{Schneider}, F.~R.~N., {Podsiadlowski}, P., \& {M{\"u}ller}, B. 2021, \aap,
  645, A5, \dodoi{10.1051/0004-6361/202039219}

\bibitem[{{Schr{\o}der} {et~al.}(2020){Schr{\o}der}, {MacLeod}, {Loeb},
  {Vigna-G{\'o}mez}, \& {Mandel}}]{Schroder20}
{Schr{\o}der}, S.~L., {MacLeod}, M., {Loeb}, A., {Vigna-G{\'o}mez}, A., \&
  {Mandel}, I. 2020, \apj, 892, 13, \dodoi{10.3847/1538-4357/ab7014}

\bibitem[{{Shigeyama} {et~al.}(1990){Shigeyama}, {Nomoto}, {Tsujimoto}, \&
  {Hashimoto}}]{1990ApJ...361L..23S}
{Shigeyama}, T., {Nomoto}, K., {Tsujimoto}, T., \& {Hashimoto}, M.-A. 1990,
  \apjl, 361, L23, \dodoi{10.1086/185818}

\bibitem[{{Shiode} \& {Quataert}(2014)}]{Shiode_Quataert_2014}
{Shiode}, J.~H., \& {Quataert}, E. 2014, \apj, 780, 96,
  \dodoi{10.1088/0004-637X/780/1/96}

\bibitem[{{Shivvers} {et~al.}(2017){Shivvers}, {Zheng}, {Van Dyk}, {Mauerhan},
  {Filippenko}, {Smith}, {Foley}, {Mazzali}, {Kamble}, {Kilpatrick},
  {Margutti}, {Yuk}, {Graham}, {Kelly}, {Andrews}, {Matheson}, {Wood-Vasey},
  {Ponder}, {Brown}, {Chevalier}, {Milisavljevic}, {Drout}, {Parrent},
  {Soderberg}, {Ashall}, {Piascik}, \& {Prentice}}]{Shivvers17}
{Shivvers}, I., {Zheng}, W., {Van Dyk}, S.~D., {et~al.} 2017, \mnras, 471,
  4381, \dodoi{10.1093/mnras/stx1885}

\bibitem[{{Smith} {et~al.}(2008){Smith}, {Foley}, \&
  {Filippenko}}]{Smith_et_al_2008}
{Smith}, N., {Foley}, R.~J., \& {Filippenko}, A.~V. 2008, \apj, 680, 568,
  \dodoi{10.1086/587860}

\bibitem[{{Smith} {et~al.}(2011){Smith}, {Li}, {Filippenko}, \&
  {Chornock}}]{Smith_et_al_2011}
{Smith}, N., {Li}, W., {Filippenko}, A.~V., \& {Chornock}, R. 2011, \mnras,
  412, 1522, \dodoi{10.1111/j.1365-2966.2011.17229.x}

\bibitem[{{Soker}(2019)}]{Soker19}
{Soker}, N. 2019, Science China Physics, Mechanics, and Astronomy, 62, 119501,
  \dodoi{10.1007/s11433-019-9402-x}

\bibitem[{{Spergel} {et~al.}(2015){Spergel}, {Gehrels}, {Baltay}, {Bennett},
  {Breckinridge}, {Donahue}, {Dressler}, {Gaudi}, {Greene}, {Guyon}, {Hirata},
  {Kalirai}, {Kasdin}, {Macintosh}, {Moos}, {Perlmutter}, {Postman},
  {Rauscher}, {Rhodes}, {Wang}, {Weinberg}, {Benford}, {Hudson}, {Jeong},
  {Mellier}, {Traub}, {Yamada}, {Capak}, {Colbert}, {Masters}, {Penny},
  {Savransky}, {Stern}, {Zimmerman}, {Barry}, {Bartusek}, {Carpenter}, {Cheng},
  {Content}, {Dekens}, {Demers}, {Grady}, {Jackson}, {Kuan}, {Kruk}, {Melton},
  {Nemati}, {Parvin}, {Poberezhskiy}, {Peddie}, {Ruffa}, {Wallace}, {Whipple},
  {Wollack}, \& {Zhao}}]{Spergel_et_al_2015}
{Spergel}, D., {Gehrels}, N., {Baltay}, C., {et~al.} 2015, arXiv e-prints,
  arXiv:1503.03757, \dodoi{10.48550/arXiv.1503.03757}

\bibitem[{{Strotjohann} {et~al.}(2021){Strotjohann}, {Ofek}, {Gal-Yam},
  {Bruch}, {Schulze}, {Shaviv}, {Sollerman}, {Filippenko}, {Yaron}, {Fremling},
  {Nordin}, {Kool}, {Perley}, {Ho}, {Yang}, {Yao}, {Soumagnac}, {Graham},
  {Barbarino}, {Tartaglia}, {De}, {Goldstein}, {Cook}, {Brink}, {Taggart},
  {Yan}, {Lunnan}, {Kasliwal}, {Kulkarni}, {Nugent}, {Masci}, {Rosnet},
  {Adams}, {Andreoni}, {Bagdasaryan}, {Bellm}, {Burdge}, {Duev}, {Dugas},
  {Frederick}, {Goldwasser}, {Hankins}, {Irani}, {Karambelkar}, {Kupfer},
  {Liang}, {Neill}, {Porter}, {Riddle}, {Sharma}, {Short}, {Taddia},
  {Tzanidakis}, {van Roestel}, {Walters}, \& {Zhuang}}]{Strotjohann_et_al_2021}
{Strotjohann}, N.~L., {Ofek}, E.~O., {Gal-Yam}, A., {et~al.} 2021, \apj, 907,
  99, \dodoi{10.3847/1538-4357/abd032}

\bibitem[{{Suzuki} {et~al.}(2019){Suzuki}, {Maeda}, \&
  {Shigeyama}}]{Suzuki_et_al_2019}
{Suzuki}, A., {Maeda}, K., \& {Shigeyama}, T. 2019, \apj, 870, 38,
  \dodoi{10.3847/1538-4357/aaef85}

\bibitem[{{Svirski} {et~al.}(2012){Svirski}, {Nakar}, \& {Sari}}]{Svirski12}
{Svirski}, G., {Nakar}, E., \& {Sari}, R. 2012, \apj, 759, 108,
  \dodoi{10.1088/0004-637X/759/2/108}

\bibitem[{{Taddia} {et~al.}(2018){Taddia}, {Stritzinger}, {Bersten}, {Baron},
  {Burns}, {Contreras}, {Holmbo}, {Hsiao}, {Morrell}, {Phillips}, {Sollerman},
  \& {Suntzeff}}]{Taddia18}
{Taddia}, F., {Stritzinger}, M.~D., {Bersten}, M., {et~al.} 2018, \aap, 609,
  A136, \dodoi{10.1051/0004-6361/201730844}

\bibitem[{{Taddia} {et~al.}(2019){Taddia}, {Sollerman}, {Fremling},
  {Barbarino}, {Karamehmetoglu}, {Arcavi}, {Cenko}, {Filippenko}, {Gal-Yam},
  {Hiramatsu}, {Hosseinzadeh}, {Howell}, {Kulkarni}, {Laher}, {Lunnan},
  {Masci}, {Nugent}, {Nyholm}, {Perley}, {Quimby}, \& {Silverman}}]{Taddia19}
{Taddia}, F., {Sollerman}, J., {Fremling}, C., {et~al.} 2019, \aap, 621, A71,
  \dodoi{10.1051/0004-6361/201834429}

\bibitem[{{Takei} \& {Shigeyama}(2020)}]{Takei20}
{Takei}, Y., \& {Shigeyama}, T. 2020, \pasj, 72, 67,
  \dodoi{10.1093/pasj/psaa050}

\bibitem[{{Takei} {et~al.}(2023){Takei}, {Tsuna}, {Ko}, \& {Shigeyama}}]{CHIPS}
{Takei}, Y., {Tsuna}, D., {Ko}, T., \& {Shigeyama}, T. 2023, {CHIPS: Complete
  History of Interaction-Powered Supernovae}, 2,  Zenodo,
  \dodoi{10.5281/zenodo.10067651}

\bibitem[{{Takei} {et~al.}(2022){Takei}, {Tsuna}, {Kuriyama}, {Ko}, \&
  {Shigeyama}}]{Takei_et_al_2022}
{Takei}, Y., {Tsuna}, D., {Kuriyama}, N., {Ko}, T., \& {Shigeyama}, T. 2022,
  \apj, 929, 177, \dodoi{10.3847/1538-4357/ac60fe}

\bibitem[{{Tan} {et~al.}(2001){Tan}, {Matzner}, \& {McKee}}]{Tan_et_al_2001}
{Tan}, J.~C., {Matzner}, C.~D., \& {McKee}, C.~F. 2001, \apj, 551, 946,
  \dodoi{10.1086/320245}

\bibitem[{{Teffs} {et~al.}(2020){Teffs}, {Ertl}, {Mazzali}, {Hachinger}, \&
  {Janka}}]{Teffs20}
{Teffs}, J., {Ertl}, T., {Mazzali}, P., {Hachinger}, S., \& {Janka}, H.~T.
  2020, \mnras, 499, 730, \dodoi{10.1093/mnras/staa2549}

\bibitem[{{Tsuna} {et~al.}(2020){Tsuna}, {Ishii}, {Kuriyama}, {Kashiyama}, \&
  {Shigeyama}}]{Tsuna20}
{Tsuna}, D., {Ishii}, A., {Kuriyama}, N., {Kashiyama}, K., \& {Shigeyama}, T.
  2020, \apjl, 897, L44, \dodoi{10.3847/2041-8213/aba0ac}

\bibitem[{{Tsuna} \& {Takei}(2023)}]{Tsuna_Takei_2023}
{Tsuna}, D., \& {Takei}, Y. 2023, \pasj, 75, L19, \dodoi{10.1093/pasj/psad041}

\bibitem[{{Tsuna} {et~al.}(2021){Tsuna}, {Takei}, {Kuriyama}, \&
  {Shigeyama}}]{Tsuna21b}
{Tsuna}, D., {Takei}, Y., {Kuriyama}, N., \& {Shigeyama}, T. 2021, \pasj, 73,
  1128, \dodoi{10.1093/pasj/psab063}

\bibitem[{{Tsuna} {et~al.}(2023){Tsuna}, {Takei}, \&
  {Shigeyama}}]{Tsuna_et_al_2023}
{Tsuna}, D., {Takei}, Y., \& {Shigeyama}, T. 2023, \apj, 945, 104,
  \dodoi{10.3847/1538-4357/acbbc6}

\bibitem[{{Valenti} {et~al.}(2008){Valenti}, {Benetti}, {Cappellaro}, {Patat},
  {Mazzali}, {Turatto}, {Hurley}, {Maeda}, {Gal-Yam}, {Foley}, {Filippenko},
  {Pastorello}, {Challis}, {Frontera}, {Harutyunyan}, {Iye}, {Kawabata},
  {Kirshner}, {Li}, {Lipkin}, {Matheson}, {Nomoto}, {Ofek}, {Ohyama}, {Pian},
  {Poznanski}, {Salvo}, {Sauer}, {Schmidt}, {Soderberg}, \&
  {Zampieri}}]{Valenti_et_al_2008}
{Valenti}, S., {Benetti}, S., {Cappellaro}, E., {et~al.} 2008, \mnras, 383,
  1485, \dodoi{10.1111/j.1365-2966.2007.12647.x}

\bibitem[{{van Loon} {et~al.}(2005){van Loon}, {Cioni}, {Zijlstra}, \&
  {Loup}}]{vanLoon05}
{van Loon}, J.~T., {Cioni}, M. R.~L., {Zijlstra}, A.~A., \& {Loup}, C. 2005,
  \aap, 438, 273, \dodoi{10.1051/0004-6361:20042555}

\bibitem[{{van Marle} {et~al.}(2010){van Marle}, {Smith}, {Owocki}, \& {van
  Veelen}}]{van_Marle_et_al_2010}
{van Marle}, A.~J., {Smith}, N., {Owocki}, S.~P., \& {van Veelen}, B. 2010,
  \mnras, 407, 2305, \dodoi{10.1111/j.1365-2966.2010.16851.x}

\bibitem[{Virtanen {et~al.}(2020)Virtanen, Gommers, Oliphant, Haberland, Reddy,
  Cournapeau, Burovski, Peterson, Weckesser, Bright, {van der Walt}, Brett,
  Wilson, Millman, Mayorov, Nelson, Jones, Kern, Larson, Carey, Polat, Feng,
  Moore, {VanderPlas}, Laxalde, Perktold, Cimrman, Henriksen, Quintero, Harris,
  Archibald, Ribeiro, Pedregosa, {van Mulbregt}, \& {SciPy 1.0
  Contributors}}]{2020SciPy-NMeth}
Virtanen, P., Gommers, R., Oliphant, T.~E., {et~al.} 2020, Nature Methods, 17,
  261, \dodoi{10.1038/s41592-019-0686-2}

\bibitem[{{Wang} {et~al.}(2023){Wang}, {Goel}, {Dessart}, {Fox}, {Shahbandeh},
  {Rest}, {Rest}, {Groh}, {Allan}, {Fransson}, {Smith}, {Hosseinzadeh},
  {Filippenko}, {Andrews}, {Bostroem}, {Brink}, {Brown}, {Burke}, {Chevalier},
  {Clayton}, {Dai}, {Davis}, {Foley}, {Gomez}, {Harris}, {Hiramatsu}, {Howell},
  {Jennings}, {Jha}, {Kasliwal}, {Kelly}, {Kool}, {Liu}, {Ma}, {McCully},
  {Miller}, {Murakami}, {Pellegrino}, {Padilla Gonzalez}, {Perera}, {Pierel},
  {Rojas-Bravo}, {Siebert}, {Sollerman}, {Szalai}, {Tinyanont}, {Van Dyk},
  {Zheng}, {Chambers}, {Coulter}, {de Boer}, {Earl}, {Farias}, {Gall},
  {McGill}, {Ransome}, {Taggart}, \& {Villar}}]{Wang_et_al_2023}
{Wang}, Q., {Goel}, A., {Dessart}, L., {et~al.} 2023, arXiv e-prints,
  arXiv:2305.05015, \dodoi{10.48550/arXiv.2305.05015}

\bibitem[{{Wheeler} {et~al.}(2015){Wheeler}, {Johnson}, \&
  {Clocchiatti}}]{Wheeler15}
{Wheeler}, J.~C., {Johnson}, V., \& {Clocchiatti}, A. 2015, \mnras, 450, 1295,
  \dodoi{10.1093/mnras/stv650}

\bibitem[{{Williamson} {et~al.}(2021){Williamson}, {Kerzendorf}, \&
  {Modjaz}}]{Williamson21}
{Williamson}, M., {Kerzendorf}, W., \& {Modjaz}, M. 2021, \apj, 908, 150,
  \dodoi{10.3847/1538-4357/abd244}

\bibitem[{{Woosley}(2017)}]{Woosley_2017}
{Woosley}, S.~E. 2017, \apj, 836, 244, \dodoi{10.3847/1538-4357/836/2/244}

\bibitem[{{Woosley}(2019)}]{Woosley2019}
---. 2019, \apj, 878, 49, \dodoi{10.3847/1538-4357/ab1b41}

\bibitem[{{Wu} \& {Fuller}(2022{\natexlab{a}})}]{Wu_Fuller_2022}
{Wu}, S.~C., \& {Fuller}, J. 2022{\natexlab{a}}, \apj, 930, 119,
  \dodoi{10.3847/1538-4357/ac660c}

\bibitem[{{Wu} \& {Fuller}(2022{\natexlab{b}})}]{Wu22_Ibc}
---. 2022{\natexlab{b}}, \apjl, 940, L27, \dodoi{10.3847/2041-8213/ac9b3d}

\bibitem[{{Yaron} {et~al.}(2017){Yaron}, {Perley}, {Gal-Yam}, {Groh}, {Horesh},
  {Ofek}, {Kulkarni}, {Sollerman}, {Fransson}, {Rubin}, {Szabo}, {Sapir},
  {Taddia}, {Cenko}, {Valenti}, {Arcavi}, {Howell}, {Kasliwal}, {Vreeswijk},
  {Khazov}, {Fox}, {Cao}, {Gnat}, {Kelly}, {Nugent}, {Filippenko}, {Laher},
  {Wozniak}, {Lee}, {Rebbapragada}, {Maguire}, {Sullivan}, \&
  {Soumagnac}}]{Yaron_2017}
{Yaron}, O., {Perley}, D.~A., {Gal-Yam}, A., {et~al.} 2017, Nature Physics, 13,
  510, \dodoi{10.1038/nphys4025}

\bibitem[{{Yoon} {et~al.}(2010){Yoon}, {Woosley}, \& {Langer}}]{Yoon10}
{Yoon}, S.~C., {Woosley}, S.~E., \& {Langer}, N. 2010, \apj, 725, 940,
  \dodoi{10.1088/0004-637X/725/1/940}

\bibitem[{{Yoshida} {et~al.}(2016){Yoshida}, {Umeda}, {Maeda}, \&
  {Ishii}}]{Yoshida16}
{Yoshida}, T., {Umeda}, H., {Maeda}, K., \& {Ishii}, T. 2016, \mnras, 457, 351,
  \dodoi{10.1093/mnras/stv3002}

\end{thebibliography}
\bibliographystyle{aasjournal}

%% This command is needed to show the entire author+affiliation list when
%% the collaboration and author truncation commands are used.  It has to
%% go at the end of the manuscript.
%\allauthors

%% Include this line if you are using the \added, \replaced, \deleted
%% commands to see a summary list of all changes at the end of the article.
%\listofchanges

\end{document}